\documentclass[10pt]{article}

\usepackage{fullpage,parskip}
\usepackage{commath}

\usepackage{wrapfig}
\usepackage[utf8]{inputenc} 
\usepackage[T1]{fontenc}    
\usepackage{hyperref,xcolor}       
\usepackage{url}            
\usepackage{booktabs}       
\usepackage{amsfonts}       
\usepackage{nicefrac}       
\usepackage{microtype}      
\usepackage{xcolor}         
\usepackage{amsmath,amsthm}
\usepackage[capitalize,noabbrev]{cleveref}
\usepackage{graphicx,amssymb}
\usepackage{algorithm,algorithmic}
\usepackage{url}
\newcommand{\email}[1]{\href{mailto:#1}{#1}}
\newtheorem{theorem}{Theorem}[section]

\newtheorem{corollary}[theorem]{Corollary}

\usepackage{multirow}
\DeclareMathAlphabet{\pazocal}{OMS}{zplm}{m}{n}

\definecolor{darkgreen}{rgb}{0.0, 0.5, 0.0}

\usepackage{amsmath}
\newcommand{\probP}{\text{I\kern-0.15em P}}
\DeclareMathOperator{\E}{\mathbb{E}}
\usepackage{xspace}
\usepackage{bbm}
\usepackage{subcaption}
\usepackage{bm}
\def\diag{\mathop{\rm diag}\nolimits}%
\newcommand{\Cc}{\mathcal{C}}
\newcommand{\Dc}{\mathcal{D}}
\newcommand{\Ec}{\mathcal{E}}
\newcommand{\Fc}{\mathcal{F}}

\newcommand{\Lc}{\mathcal{L}}

\newcommand{\Qc}{\mathcal{Q}}

\newcommand{\Sc}{\mathcal{S}}

\newcommand{\Xc}{\mathcal{X}}
\newcommand{\Yc}{\mathcal{Y}}

\newcommand{\ex}{{\rm e}}

\newcommand{\ev}{{\bf e}}

\newcommand{\Av}{{\bf A}}
\newcommand{\Bv}{{\bf B}}
\newcommand{\Cv}{{\bf C}}
\newcommand{\Dv}{{\bf D}}
\newcommand{\Hv}{{\bf H}}
\newcommand{\Xv}{{\bf X}}
\newcommand{\Yv}{{\bf Y}}
\newcommand{\Zv}{{\bf Z}}

\newcommand{\xv}{{\bf x}}
\newcommand{\yv}{{\bf y}}
\newcommand{\zv}{{\bf z}}

\newcommand{\sv}{{\bf s}}
\newcommand{\cv}{{\bf c}}
\newcommand{\dv}{{\bf d}}

\newcommand{\muv}{\boldsymbol \mu}
\newcommand{\thetav}{\bm{\theta}}

\newcommand{\xhv}{{\bf \hat{x}}}
\newcommand{\xbv}{{\bf \bar{x}}}
\newcommand{\xbhv}{\bar{\hat{\xv}}}
\newcommand{\xtv}{{\bf \tilde{x}}}

\newcommand{\ts}{^{\top}}

\newcommand{\bthe}{{\Bar{\theta}}}
\newcommand{\bthev}{{\Bar{\thetav}}}

\newcommand{\ie}{{\em i.e.}}

\newcommand{\mvec}[1]{{\boldsymbol #1}}
\newcommand{\Lp}[3]{{\left\|{#1}\right\|}_{#2}^{#3}}

\title{Theoretical Characterization of Effect of Masks in Snapshot Compressive Imaging}

\author{
    Mengyu Zhao$^*$, Shirin Jalali\thanks{Rutgers University, New Brunswick, Department of Electrical and Computer Engineering (\email{mengyu.zhao@rutgers.edu}, \email{shirin.jalali@rutgers.edu}).}\\
}
\date{}
\begin{document}

\maketitle

\begin{abstract}
Snapshot compressive imaging (SCI) refers to the recovery of three-dimensional data cubes—such as videos or hyperspectral images—from their two-dimensional projections, which are generated by a special encoding of the data with a mask.   SCI systems commonly use binary-valued masks that follow certain physical constraints. Optimizing  these masks subject to these constraints is expected to improve system performance. However, prior theoretical work on SCI systems focuses solely on independently and identically distributed (i.i.d.) Gaussian masks, which do not permit such optimization. On the other hand, existing practical mask optimizations rely on computationally intensive  joint optimizations that provide limited insight into the role of masks and are expected to be sub-optimal due to the non-convexity and complexity of the optimization. In this paper, we analytically characterize the performance of SCI systems employing binary masks and leverage our analysis to optimize hardware parameters. Our findings provide a comprehensive and fundamental understanding of the role of binary masks — with both independent and dependent elements — and their optimization. We also present simulation results that confirm our theoretical findings and further illuminate different aspects of mask design.
\end{abstract}


\section{Introduction}
Snapshot compressive imaging (SCI) systems optically encode a high-dimensional (HD) 3D data cube, such as a video or a hyperspectral image, into a 2D image. The SCI recovery algorithm then reconstructs  the desired 3D data cube from the 2D measurements.  In recent years, SCI optical encoding solutions have been proposed for various types of data. (For an overview, refer to  \cite{yuan2021snapshot}.) The primary goal of such solutions is to improve the efficiency of the data acquisition process.  For instance, a key application of SCI is in hyperspectral imaging (HSI), an emerging technology with a wide range of applications,  from medicine to astronomy, see e.g. \cite{liu2011tongue,lu2014medical,hege2004hyperspectral,kurz2013close,peyghambari2021hyperspectral}. Standard  HSI solutions rely on scanning the data either in space or along the wavelengths, which makes the process slow and costly. SCI solutions for hyperspectral imaging, on the other hand,  dramatically speed up the process by capturing all the information in a single snapshot \cite{gehm2007single,rajwade2013coded}.

In  SCI systems, the \emph{masks}, used for optical encoding of data, plays a critical role by enabling reconstruction of the 3D data cube from the 2D projection. While such (hardware) masks can be controlled and designed in practice, they are typically subject to various constraints.  For instance, these masks are typically binary-valued. Additionally, the masks used for different frames are often dependent. In some cases, they might be randomly shifted versions of each other, as described in \cite{llull2013coded}. This raises the following question:

     \noindent{\bf Question}: Can we theoretically design the SCI masks subject to the limitations to optimize the recovery performance? 

Mathematically, the SCI optical encoding  can be modeled as a highly under-determined linear inverse problem.  Over the past decade, compressed sensing - the technique of reconstructing structured signals from an ill-posed linear inverse problem – has been a well-researched area in the scientific literature. Numerous works have focused on designing algorithms for general reconstruction tasks, including both model-based and learning-based approaches; see, e.g., \cite{liao2014generalized, Pendupnpimagerestoration2023, Koflertvunrolling2023, Goujonconvexregu2024,caiimpainting2024}, offering valuable insights.  However, the specialized structure of SCI sensing matrices on one hand, and the complex structure of the input data on the other hand, limits the  applicability of compressed sensing results to these systems. Prior work \cite{jalali2019snapshot} demonstrated the theoretical feasibility of SCI recovery under the assumption of independent and identically distributed (iid) Gaussian masks, but such models deviate significantly from real-world SCI systems.

This paper aims to bridge this gap by analyzing SCI systems under more realistic mask models. Specifically, we focus on binary-valued masks with dependencies across frames, reflecting the constraints of practical SCI systems. Through this analysis, we seek to achieve two objectives: first, to identify the characteristics of optimal masks, and second, to understand how the limitations of binary-valued and dependent masks impact achievable performance.

To achieve this goal, we adopt the compression-based framework initially proposed in \cite{jalali2016compression} and later utilized in \cite{jalali2019snapshot} for the theoretical analysis of SCI systems. Compression-based methods are known to achieve the fundamental limits in compressed sensing, in settings where these limits are known \cite{RezagahJ:17-IT}. They provide a versatile framework for studying inverse problems without explicitly specifying the source model. We theoretically characterize an upper bound on the achievable error for an idealized compression-based SCI recovery under the following models for the elements of the masks:
\begin{enumerate}
    \item {\bf Fully independent masks.} Masks corresponding to different frames are designed independently, and the elements of each mask are independently and identically distributed (i.i.d.).
    \item {\bf In-frame dependent masks.} Masks corresponding to different frames are designed independently, but the elements within each frame are dependent. We mathematically model these dependencies as a binary Markov process within each frame.
    \item {\bf Out-of-frame dependent masks.} Masks corresponding to different frames are dependent, such that elements encoding collocated pixels across the frames are modeled as a binary Markov process. Otherwise, the elements of the masks are independent.
\end{enumerate}
Our theoretical contributions can be summarized as follows: 
\begin{enumerate}
    \item In the case of noise-free measurements, we show that the error bound is minimized when the elements of the masks are fully independent and follow a Bernoulli distribution ${\rm Bern}(p)$ with $p^*<0.5$. This result is consistent with empirical observations reported in the literature \cite{iliadis2020deepbinarymask, zhang2022herosnet, zhao2024untrained}. We also show that dependence—whether in-frame or out-of-frame—inevitably degrades  performance. 
    \item We prove that the projected gradient descent (PGD) algorithm can effectively solve the idealized compression-based optimization problem. Specifically, we prove that the PGD algorithm, with an appropriately chosen step size, converges to the vicinity of the optimal solution despite the problem's non-convex nature. This guarantees the practical viability of compression-based methods in achieving near-optimal performance.
    \item To validate our theoretical findings, we conduct extensive experiments on video SCI using the PGD algorithm with two projection methods: a traditional denoiser (GAP-TV \cite{yuan2016generalized}) and a pretrained deep denoiser (PnP-FastDVDnet \cite{PnP_fastdvd}).
  
    \begin{enumerate}
        
\item Our empirical results show that optimized mask properties for reconstruction with binary i.i.d. masks align with theoretical predictions.
\item We investigate the effect of varying the number of video frames used during mask optimization, as well as the impact of both in-frame and out-of-frame dependencies. Simulation results consistently validate that any form of dependency degrades performance.
\end{enumerate}
\end{enumerate}



\subsection{Related work}

Optimizing measurement strategies and masks in SCI  has been the subject of various studies in the recent literature.  In \cite{iliadis2020deepbinarymask}, the authors studied training the binary masks, and observed that the empirically optimized masks exhibit smooth variations and a nonzero element probability around 40\% (60\% sparsity). Deep unfolding networks have also been employed to tackle the challenge of simultaneous image reconstruction and mask optimization in SCI, resulting in preserved image structure and optimal sampling patterns~\cite{zhang2022herosnet}. The benefits of mask optimization are further supported by hardware prototypes comparing random and optimized masks~\cite{koller2015high}. Expanding on these concepts, researchers have explored incorporating apertures for mask multiplexing in SCI, offering additional performance gains~\cite{Zhang_2021}. Furthermore, end-to-end network architectures have been proposed for joint mask and network optimization in SCI, leading to improvements in loss function and peak signal-to-noise ratio (PSNR)~\cite{9105237}.  Finally, optimizing the masks' parameters has also been studied for unsupervised SCI recovery solutions, which employ untrained neural networks, such as deep image prior \cite{ulyanov2018deep}, to model the source structure \cite{zhao2024untrained}. In summary, these studies highlight the  impact of mask  optimization on enhancing SCI system performance.

\subsection{Notations}
Vectors are denoted by bold letters, such as $\xv$ and $\yv$. For a matrix $\Xv\in\mathbb{R}^{n_1\times n_2}$, ${\rm Vec}(\Xv)$ denotes the  vector in $\mathbb{R}^n$, $n=n_1 n_2$, formed by concatenating the columns of $\Xv$. $\odot$ denotes the Hadamard matrix product operator. That is, for $\Av,\Bv\in\mathbb{R}^{n_1\times n_2}$, $\Yv=\Av\odot \Bv$ is defined such that $Y_{ij}=A_{ij}B_{ij}$, for all $i,j$. Sets are denoted by calligraphic letters, such as $\Xc,\Yc$. For a finite set $\Xc$, $|\Xc|$ denotes the size of $\Xc$

\subsection{Outline}
 Section \ref{sec:SCI} defines the SCI mathematical problem statement. Section \ref{sec:CSP} reviews compression-based methods for solving SCI inverse problems as the theoretical framework adopted in this paper. Sections \ref{sec:main} and \ref{sec:pgd} present the main theoretical results of the paper on characterizing the performance of SCI systems under different mask distributions, for idealized compression-based optimization and the corresponding PGD, respectively.  Section \ref{sec:experiments} presents  our numerical results confirming our theoretical findings.  Proofs of the mains results are presented in Section \ref{sec:proofs}.  Section \ref{sec:conclusion} concludes the paper. 

\section{Problem statement}\label{sec:SCI}
The goal of an SCI system is to recover a 3D data cube from its 2D projection, while knowing the mapping. More precisely,  let $\Xv \in \mathbb{R}^{n_1 \times n_2 \times B}$ denote the desired 3D data cube. An SCI system maps $\Xv$ to a single measurement frame  $\Yv \in \mathbb{R}^{n_1\times n_2}$. In many SCI systems, such as HS SCI \cite{gehm2007single} and video SCI \cite{llull2013coded},  the mapping from $\Xv$ to $\Yv$ can be modeled as a linear system such that \cite{llull2013coded,Wagadarikar09CASSI}, $\Yv = \sum_{b=1}^B \Cv_b\odot \Xv_b + \Zv$, where $\Cv\in \mathbb{R}^{n_1 \times n_2 \times B}$  and $\Zv \in \mathbb{R}^{n_1 \times n_2 }$ denote the sensing kernel  (mask) and the additive noise, respectively. Here, $\Cv_b = \Cv(:,:,b)$ and $\Xv_b = \Xv(:,:,b) \in \mathbb{R}^{n_1 \times n_2}$ represent the $b$-th sensing kernel (mask) and the corresponding signal frame, respectively.

\begin{figure}[h]
\begin{centering}
\includegraphics[width=8cm]{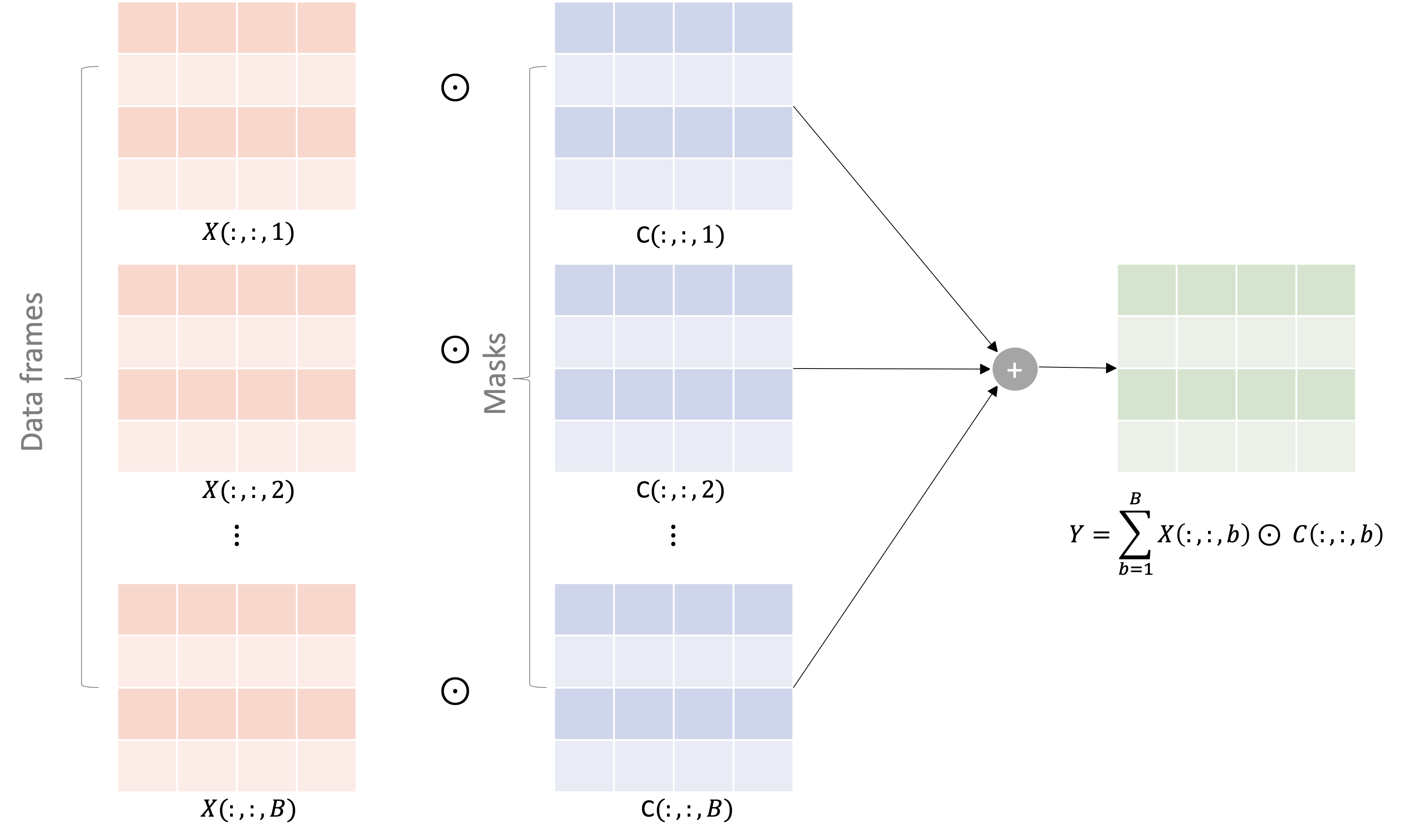}
\par\end{centering}
\caption{SCI encoding function: For $b=1,\ldots,B$, frame $b$ and mask $b$ are represented  by  $X(:,:,b)$ and $C(:,:,b)$, respectively. The single  2D measurement frame is generated as $\sum_{b=1}^BX(:,:,b)\odot C(:,:,b)$.}\label{fig:sys-model}
\end{figure}

To simplify the mathematical representation of the system, we vectorize each frame as  $\xv_b={\rm Vec}(\Xv_b)\in\mathbb{R}^n$ with $n = n_1 n_2$.  Then, we vectorize the data cube  $\Xv$ by concatenating the $B$ vectorized frames into a column vector $\xv \in \mathbb{R}^{n B}$ as
\begin{equation} \label{Eq:xv1toB} 
\xv = \left[\begin{array}{c}
\xv_1\\
\vdots\\
\xv_B
\end{array}\right].
\end{equation}
Similarly, we define  $\yv = \text{Vec}(\Yv) \in \mathbb{R}^{n}$ and $\zv= \text{Vec}(\Zv) \in \mathbb{R}^{n}$. Using these definitions, the measurement process defined in Fig.~\ref{fig:sys-model} can also be expressed  as 
\begin{align}
\yv = \Hv \xv + \zv.\label{eq:SCI-model}
\end{align}
The sensing matrix $\Hv\in\mathbb{R}^{n\times nB}$, is a highly sparse matrix that is formed by the  concatenation of $B$  diagonal matrices as
\begin{equation}\label{Eq:Hmat_strucutre}
\Hv = [\Dv_1,...,\Dv_B],
\end{equation}
where, for  $b =1,\dots B$, $\Dv_b = \text{diag}(\text{Vec}(\Cv_b)) \in {\mathbb R}^{n \times n}$. 
The goal of a SCI recovery algorithm is to recover the data cube $\xv$ from undersampled measurements $\yv$, while having access to the sensing matrix (or mask) $\Hv$.

\section{Compression-based SCI recovery}\label{sec:CSP}


Consider the problems of SCI defined in Section \ref{sec:CSP}. To recover $\xv$ from underdetermined measurements $\yv$, we need to take advantage of the structure of $\xv$. However,  the desired mathematical model of the structure needs to capture both intra- or inter-frame dependencies, which makes designing such mathematical models inherently  complex. One approach to address this issue and provide a theoretical analysis is SCI systems is to adopt the idea of compression-based recovery \cite{jalali2016compression,RezagahJ:17-IT,jalali2016compression}. The key advantage of this approach is that instead of explicitly expressing the structure, it will be captured through a compression  code, and the performance is determined by the key parameters of the compression code, i.e., its rate $r$ and distortion $\delta$.

Consider a compact set $\Qc\subset\mathbb{R}^{nB}$.  Each signal $\xv\in\Qc$,
consists of $B$ vectors (frames) $\{\xv_1\ldots\xv_B\}$ in $\mathbb{R}^n$.  A lossy compression code of rate $r$ for $\Qc$  is characterized by its encoding mapping $f$, where $f: \Qc \rightarrow \{1,2,\ldots, 2^{Br}\},$ and $g: \{1,2\ldots2^{Br}\}\to \mathbb{R}^{nB}$.
For $\xv\in\Qc$,  $\tilde{\xv} = g(f(\xv))$ denotes the reconstruction corresponding to $\xv$. The {\emph {  distortion}} between  $\xv$ and its reconstruction  $\hat{\xv}$ is defined as 
\begin{equation}
 d(\xv,\hat{\xv})\triangleq   {1\over nB}\|\xv-\hat{\xv}\|_2^2.
 \label{eq:def-delta-B}
\end{equation}
The compression code $(f,g)$ is characterized by its rate $r$ and distortion  $\delta$ defined as 
\[
\delta=\sup_{\xv\in\Qc}  d(\xv,g(f(\xv))).
\]
Moreover, the defined encoder and decoder pair, $(f,g)$, corresponds to a codebook  $\Cc$ defined as
\begin{equation}
\Cc=\{g(f(\xv)):\; \xv\in\Qc\}.  \label{eq:codebook}
\end{equation}
Note that $|\Cc|\leq 2^{Br}$.

Given a class of signal denoted by a $\Qc\subset\mathbb{R}^{nB}$, and a rate-$r$ distortion-$\delta$  compression code $(f,g)$, the compressible signal pursuit (CSP) optimization recovers $\xv\in\Qc$ from measurements $\yv\in\mathbb{R}^n$ defined in \eqref{eq:SCI-model}, as follows
\begin{align}
\hat{\xv}=\arg\min_{\cv\in \Cc}\|\yv-\sum_{i=1}^B\Dv_i\cv_i\|_2^2.\label{eq:CSP}
\end{align}
The performance of \eqref{eq:CSP} is theoretically characterized in \cite{jalali2019snapshot} for the case where the diagonal  entries of $\Dv_1,\ldots,\Dv_B$ are i.i.d.~Gaussian. However, the practical masks used in real-world applications deviate from this i.i.d. Gaussian model in several ways: 1) They are typically binary-valued, and 2) In some applications, the elements of masks used for different frames are dependent. In this paper, we consider mathematical models that align with practical masks and perform a theoretical analysis of SCI recovery algorithms. Our goal is to provide a theoretical foundation for optimizing the masks and to understand the impact of dependencies on the performance.



\section{Theoretical results on effect of masks}
\label{sec:main}

In this section, we present our main theoretical results on the performance of SCI systems under different settings of binary masks. Our goal is to address the questions we raised before on how the statistical properties and dependencies of binary masks impact the performance of SCI systems, and whether it is possible to optimize the masks to achieve better performance. We discuss our findings in three distinct settings, each corresponding to different mask characteristics and their effects on the system's performance.

 Let $\Qc\subset\mathbb{R}^{nB}$ denote the set of desired signals, for example vectorized videos consisting of $B$ frames. For all $\xv\in\Qc$, assume that $\|\xv\|_{\infty}\leq\frac{\rho}{2}$.  Consider a rate-$r$ distortion-$\delta$ lossy compression code for signals in $\Qc$ and let $\Cc$ denote the codebook corresponding to this compression code. For $\xv\in\Qc$, the SCI encoder measures  $\yv=\sum_{i=1}^B\Dv_i\xv_i$, where as defined earlier, for $i=1,\ldots,B$,  $\Dv_i={\rm diag}(D_{i1}\ldots D_{in})$. In the rest of this section, we theoretically analyze the performance of a compression-based SCI recovery algorithm, under different distributions on the elements of $\Dv_1,\ldots,\Dc_B$.

\subsection{i.i.d.~Bernoulli masks}
As the first scenario, we focus on the masks entries being i.i.d.~ and binary-valued. There are two key questions we want to address in this setting: Is recovery still feasible? If so, what is the optimal value of $p$,  $p=\Pr(D_{ij}=1)$,  that minimizes the achieved distortion between the signal and its SCI reconstruction?

\begin{theorem}\label{thm:1} 
For $\xv\in\Qc$ and $\yv=\sum_{i=1}^B\Dv_i\xv_i$ let $\hat{\xv}$ denote the solution of \eqref{eq:CSP}. We assume that $\Dv_1,\ldots,\Dv_B$ are  independent and, for $i=1,\ldots,B$, $(D_{i1}\ldots D_{in})$ are   i.i.d.~${\rm Bern}(p)$. Choose free parameter $\eta>0$. 
Let 
\[
\xbhv={1 \over B}\sum_{i=1}^B\hat{\xv}_i.
\]
Then,
\begin{equation}
\frac{1}{nB}\|\xv-\hat{\xv}\|_2^2+\big({p\over 1-p}\big)\big(\frac{1}{n}\|\xbv-\xbhv\|_2^2\big)\leq(1+\frac{Bp}{1-p})\delta+{\rho^2B\over (p-p^2)}\sqrt{(B+\eta){2r\over n}},\label{eq:main-thm1-result}
\end{equation}
with a probability larger than $1-2\exp(-\eta r)$. Moreover, for fixed parameters $(n,B,\rho,\eta)$, the bound in \eqref{eq:main-thm1-result} is minimized at some $p^*$, where $p^*<{1\over 2}$. 
\end{theorem}

The following corollary of Theorem \ref{thm:1} characterizes the number of frames $B$ that can be recovered from a 2D SCI measurement, with high probability, using \eqref{eq:CSP}.

\begin{corollary}\label{cor:4-2}
    Given $\kappa\geq 1$, 
    \[
    B\leq ({\kappa\delta\over \rho^2})^{2\over 3}({n\over r})^{1\over 3},
    \]
    then, with a probability larger than $1-\exp(- r)$, we have 
    \begin{equation}
\frac{1}{nB}\|\xv-\hat{\xv}\|_2^2\leq\Big(1+\frac{Bp}{1-p}+{2\kappa \over (p-p^2)}\Big)\delta.
\end{equation}
 
\end{corollary}

It can be observed that  as $p$ approaches zero or one,  the bound in \eqref{eq:main-thm1-result} grows without bound, confirming that one cannot  expect recovery from an all-zero or an all-one mask. On the other hand, for $p=0.5$, Theorem \ref{thm:1} guarantees that 
\[
\frac{1}{nB}\|\xv-\hat{\xv}\|_2^2\leq(1+B)\delta+{4\rho^2}B^2\sqrt{(1+\eta){2r\over n}},
\]
with probability larger than $1-\exp(-\eta r)$. Moreover, it states that the optimal $p^*$ is smaller than $0.5$, which means that the optimal bound is tighter than this result. This is consistent with the results from the literature, e.g.~\cite{iliadis2020deepbinarymask}, that show through various types of algorithmic optimizations that in the learned optimized binary masks  $\Pr(D_{ij}=1)$ is strictly smaller than $0.5$. 

One distinctive property of binary masks compared to i.i.d.~Gaussian masks studied in the prior art is that $D_{i,j}\geq 0$, w.p. $1$. To further highlight this difference and show its potential impact on optimizing the masks, in the following corollary of Theorem \ref{thm:1}, we consider the case where instead of being binary-valued, the masks take values in $\{-1,+1\}$. Under this model, the optimal bound on the distortion  is achieved at $p^*=0.5$, which implies that under the optimal setting $\E[D_{i,j}]=0$.

\begin{corollary}\label{cor:1} Consider the same setup as in Theorem 1, where instead of binary masks,  $D_{ij}\in\{-1,+1\}$ and $\{\{D_{ij}\}_{j=1}^n\}_{i=1}^B$ are i.i.d.~ such that $\Pr(D_{ij}=1)=1-\Pr(D_{ij}=-1)=p$. Then
\[
\frac{1}{nB}\|\xv-\hat{\xv}\|_2^2\leq\big(1-B+{B \over 4(p-p^2)}\big)\delta+{\rho^2B\epsilon\over 4(p-p^2)},
\]
with a probability larger than $1-2^{Br+1}\exp(-n\epsilon^2/2)$.
The upper bound is minimized at $p^*={1\over 2}$, which leads to $\frac{1}{nB}\|\xv-\hat{\xv}\|_2^2\leq\delta+{\rho^2B\epsilon}$.
\end{corollary}

Throughout this section we focused on masks that either are binary-valued ($D_{i,j}\in\{0,1\}$) or $D_{i,j}\in\{-1,+1\}$ (in Corollary \ref{cor:1}). However, we can extend the result to the case where the masks can have any general distribution that is confined to a bounded interval. For instance, in the following Corollary, we study the case where the elements of the masks are restricted to $[0,1]$ interval. 
\begin{corollary}\label{cor:1.2} Consider the same setup as in Theorem \ref{thm:1}. Assume that  $\{\{D_{ij}\}_{j=1}^n\}_{i=1}^B$ are i.i.d.~ such that i) $0\leq D_{ij}\leq 1$, ii) $\E[D_{ij}]=p$ and iii) $\E[D_{ij}^2]=q$. Then, 
\begin{equation}
\frac{1}{nB}\|\xv-\hat{\xv}\|_2^2\leq(1+\frac{Bp^2}{q-p^2})\delta+{\rho^2B\epsilon\over (q-p^2)}\label{eq:bound-cor4-4}
\end{equation}
with a probability larger than $1-2^{Br+1}\exp(-n\epsilon^2/2)$.
\end{corollary}
Note that given that the $D_{i,j}$s are within $[0,1]$,
\[
p=\E[D_{ij}]\geq \E[D_{ij}^2]=q.
\]
On the other hand, it can be observed that the upper bound in \eqref{eq:bound-cor4-4} decreases monotonically as a function of \(q\). Therefore, for a fixed \(p\), the bound is minimized by maximizing \(q\). The value of \(q\) is maximized when \(D_{ij}\) is binary-valued, which implies \(p = \E[D_{ij}] = \E[D_{ij}^2] = q\). In this case, the bound simplifies to the form we presented earlier.



\subsection{Binary Markov masks: in-frame dependence}

Consider a setting where masks corresponding to different frames are independent, but the entries of each mask are dependent and follow a first-order Markov process. More precisely,  assume that $\Dv_1,\ldots,\Dv_B$ are independent. Also, for  $i=1,\ldots,B$, the diagonal entries of $\Dv_i$ are generated according to a stationary Markov process such that, for $j=2,\ldots,n$, 
\[
p_{D_{ij}|D_{i,1:(j-1)}}(\cdot|\cdot)=p_{D_{ij}|D_{i,j-1}}(\cdot|\cdot).
\]
Moreover, for any $i=1,\ldots,B$, and $j=2,\ldots,n$, we define the transition kernel of the (asymmetric) Markov chain as follows
\begin{align}
    \Pr(D_{ij}=1|D_{i(j-1)}=0)&=q_0,\nonumber\\
    \Pr(D_{ij}=0|D_{i(j-1)}=1)&=q_1.\label{eq:in-frame-dist}
\end{align}

%
%

%
%
%
%
%
%
%
%
%
%
%
%
%
%
%

%

\begin{figure}[h]
\begin{center}
\includegraphics[width=7cm]{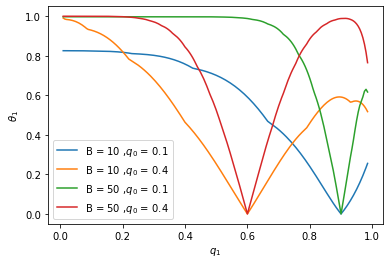}\caption{ $\theta_1$ as a function of $q_1$.}\label{fig:2}
\end{center}
\end{figure}

To characterize the performance under the described Markov model for the masks, we use the concentration of measure results developed in \cite{kontorovich2008concentration}. For using that result, we define the contraction coefficient corresponding to the defined Markov process as 
\begin{align}
    \theta_1 
    &= \sup_{\dv',\dv''\in\Sc^B}\| p(\cdot|\dv')-p(\cdot|\dv'')\|_{\rm TV}\nonumber\\
    &= \| p_i(\cdot|\dv'={\bf 0}_B)-p_i(\cdot|\dv''={\bf 1}_B)\|_{\rm TV}\nonumber\\
    &= \frac{1}{2}\Big[| q_0^B-(1-q_1)^B|   + \binom{B}{1}|(1-q_0)q_0^{B-1}-q_1(1-q_1)^{B-1}|\nonumber\\ 
       &\;\;+ \binom{B}{2}|(1-q_0)^2q_0^{B-2}-q_1^2(1-q_1)^{B-2}|  + \cdots+ | (1-q_0)^B-q_1^B|\Big].\label{eq:def-theta1}
\end{align}
 In \eqref{eq:def-theta1}, for $\dv,\dv'\in\Sc^B$, $p(\dv'|\dv)=\prod_{i=1}^Bp(d'_i|d_i)$ denotes the transition kernel of the defined Markov process. Fig.~\ref{fig:2} shows the value of $\theta_1$ as a function of $q_1$, for a couple of different  values of $q_0$ and $B$. 

\begin{theorem}\label{thm:2}
 Assume that $(D_{i1},\ldots,D_{in})$, $i=1,\ldots,B$, are independently generated as stationary first order Markov processes with transition probabilities described in \eqref{eq:in-frame-dist}. 
For $\xv\in\Qc$ and $\yv=\sum_{i=1}^B\Dv_i\xv_i$ let $\hat{\xv}$ denote the solution of \eqref{eq:CSP}. 
Then
\begin{equation}
\frac{1}{nB}\|\xv-\hat{\xv}\|_2^2\leq(1+\frac{Bp}{1-p})({\delta\over nB})+{\rho^2\epsilon\over p(1-p)},\label{eq:main-thm2-bd}
\end{equation}
with a probability larger than 
\[
1-(2^{Br}+1)\exp(-\frac{n\epsilon^2}{32}(1-\theta_1)^2),
\]
 where $\theta_1$ is defined in \eqref{eq:def-theta1}.
\end{theorem}

Comparing the bound in \eqref{eq:main-thm2-bd} and the one in \eqref{eq:main-thm1-result} shows that they are indeed equivalent. Therefore, similar to Theorem \ref{thm:1}, the bound is minimized for some $p^*={q_0^*\over q_0^*+q_1^*}<0.5$. On the other hand, to minimize $\exp(-\frac{n\epsilon^2}{32}(1-\theta_1)^2)$ which controls how many frames can be decoupled from each other, we need to minimize $\theta_1$ defined in \eqref{eq:def-theta1}. But we can set $\theta_1=0$, by setting $q_0^*=p^*$ and $q_1^*=1-p^*$.  It is straightforward to see that setting the parameters  $q_0$ and $q_1$ as such corresponds to making the Markov process an independent process. This is intuitively not surprising as using this setting the convergence speed of the random variables is maximized.


\subsection{Binary Markov masks: Out-of-frame dependence}
Next we consider the case where the entries of each mask are generated independently, but the mask entries corresponding to element $i$ of each frame are dependent. This is closely related to real masks used in practice where each mask can be a shifted version of the previous mask. Mathematically, we assume that $D_{1j},\ldots,D_{Bj}$ are generated according to a stationary first order Markov process such that for any $j=1,\ldots,n$ and $i=2,\ldots,B$,
\[
p_{D_{ij}|D_{1:(i-1),j}}(\cdot|\cdot)=p_{D_{ij}|D_{i-1,j}}(\cdot|\cdot),
\]
and 
\begin{align}
\Pr(D_{ij}=1|D_{i-1,j}=0)&=q_0\nonumber\\
\Pr(D_{ij}=0|D_{i-1,j}=1)&=q_1.\label{eq:out-frame-dist}
\end{align}
Assume that $q_0,q_1\leq 0.5$ and let 
\[
\alpha=1-q_0-q_1.
\]
Note that since $q_0,q_1\leq 0.5$, $\alpha\geq 0$. 
Define $B\times B$ matrix $\Lambda$ as follows
\begin{align}
\Lambda =
\begin{bmatrix}
1 & \alpha & \cdots & \alpha^{B-1} \\
    \alpha & 1& \cdots & \alpha^{B-2} \\
    \vdots & \vdots & \ddots & \vdots \\
    \alpha^{B-1} & \alpha^{B-2} & \cdots &1.
\end{bmatrix}.\label{eq:def-Lambda}
\end{align}
Let $\lambda_{\max}(\Lambda)$ and $\lambda_{\min}(\Lambda)$ denote the maximum and minimum eigenvalues of matrix $\Lambda$, respectively. 

\begin{theorem}\label{thm:3}
 Assume  that  $\Dv_1\ldots\Dv_B$  are such that $\Dv_i={\rm diag}(D_{i1}\ldots D_{in})$, $i=1,\ldots,B$, where $D_{ij}\in\{0,1\}$. Assume that $(D_{1j},\ldots,D_{Bj})$, $j=1,\ldots,n$, are independently generated as stationary first order Markov processes according to \eqref{eq:out-frame-dist}. For $\xv\in\Qc$ and $\yv=\sum_{i=1}^B\Dv_i\xv_i$ let $\hat{\xv}$ denote the solution of \eqref{eq:CSP}. 
Then, if $ \lambda_{\min}(\Lambda)>0$, 
\begin{align*}
\frac{1}{nB}\|\xv-\hat{\xv}\|_2^2\leq &{\lambda_{\max}(\Lambda)(1-p) +pB \over \lambda_{\min}(\Lambda)(1-p)}({\delta\over nB})\nonumber\\
&+{\rho^2\epsilon\over \lambda_{\min}(\Lambda) p(1-p)},
\end{align*}
 with a probability larger than $1-2^{Br+1}\exp(-\frac{n\epsilon^2}{2B^2})$.
\end{theorem}
Note that in the case where all the entries of the sensing matrix are independent, i.e., the case where $q_0+q_1=1$ and $\alpha=0$,  $ \lambda_{\max}(\Lambda)= \lambda_{\min}(\Lambda)=1$. Therefore, in that case the upper bound in Theorem \ref{thm:3} simplifies to the result of Theorem \ref{thm:1}. 

We can use Gershgorin circle theorem \cite{gershgorin1931uber} to derive upper and lower bounds on $\lambda_{\max}(\Lambda)$ and $\lambda_{\min}(\Lambda)$, respectively, and find the following corollary. 
\begin{corollary}\label{cor:thm-3}
Consider the same setup as in Theorem \ref{thm:3}. Then, for $\alpha<{1\over 3}$, 
\begin{align*}
\frac{1}{nB}\|\xv-\hat{\xv}\|_2^2\leq &{(1+\alpha)(1-p) +pB \over (1-3\alpha)(1-p)}({\delta\over nB})\nonumber\\
&+{\rho^2(1-\epsilon) \over(1-3\alpha) p(1-p)},
\end{align*}
 with a probability larger than $1-2^{Br+1}\exp(-\frac{n\epsilon^2}{2B^2})$.
\end{corollary}


\section{Recovery algorithm: Compression-based projected gradient descent}
\label{sec:pgd}

So far we focused on the analyzing the effect of the masks distribution on the performance of \eqref{eq:CSP}, which is a non-convex optimization. However, solving \eqref{eq:CSP} directly, due to its nature which involves minimizing a convex function over a discrete set of exponentially many codewords is computationally prohibitive. To address this challenge and design an implementable algorithm, a classic approach is to use iterative algorithms that are inspired by the Projected gradient descent (PGD).  PGD is a well-established method for solving convex optimization problems, and extensive research has explored the convergence performance of this algorithm \cite{nesterov2013introductory}. More recent results, such as \cite{attouch2013convergence}, also investigate the performance of such algorithms when applied to non-convex problems, which are different from those considered in this paper. 

Directly applying the PGD algorithm to solve \eqref{eq:CSP} leads to Algorithm~\ref{algo:PGD}.  Each iteration of Algorithm~\ref{algo:PGD} consists of two main steps: 
\begin{itemize}
	\item [i)] moving in the direction of the gradient of the cost function,
	\item [ii)] projecting the result onto the set of codewords.
\end{itemize}
Both steps remain computationally efficient. In the gradient descent step, matrix-vector multiplication is performed, \ie, $\Hv \xv^t$ and $\Hv \ts \ev^t$, with $\ev^t = \yv - \Hv \xv^t$. In this work, $\Hv = [\Dv_1, \ldots, \Dv_B]$ represents a binary matrix, rather than a matrix with Gaussian entries, reflecting a key distinction from prior work~\cite{jalali2019snapshot}. The second step, projecting onto the set of codewords, is achieved by applying the encoder and decoder of the compression code, similar to the previous Gaussian case.

\begin{algorithm}[htbp!]
	\caption{Compression-based PGD for SCI recovery}
	\begin{algorithmic}[1]
		\REQUIRE$\Hv$, $\yv$.
		\STATE Initial $\mu>0$, $\xv^0 = 0$.
		\FOR{$t=0$ to Max-Iter }
		\STATE Calculate: $\ev^t = \yv - \Hv \xv^t$.
		\STATE Projected gradient descent:  $\sv^{t+1} = \xv^{t} + \mu \Hv\ts \ev^t$.
		\STATE Projection via compression:  $\xv^{t+1} = g(f(\sv^{t+1}))$.
		\ENDFOR
		\STATE $\textbf{Output:}$ Reconstructed signal $\hat \xv$.
	\end{algorithmic}
	\label{algo:PGD}
\end{algorithm}

Our next theorem, shows that when the entries of the masks are i.i.d.~${\rm Bern}(p)$, setting the step size as $\mu=1/(p-p^2)$, if the number of frames $B$ is small enough, then the PGD algorithm converges to the vicinity of the desired result.

\begin{theorem}\label{thm:pgd}
Consider a compact set ${\cal Q}\subset\mathbb{R}^{nB}$, such that for all $\mvec{x}\in{\cal Q}$, $\Lp{\mvec{x}}{\infty}{}\leq {\rho\over 2}$. Furthermore, consider a compression code for set ${\cal Q}$ with encoding and decoding mappings, $f$ and $g$, respectively.  Assume that the code  operates at rate $r$ and distortion $\delta$. Let $\tilde{\mvec{x}}=g(f(\mvec{x}))$. Assume that $\mvec{x}$ is measured as $\mvec{y}=\sum_{i=1}^B\Dv_i\mvec{x}_i$, where $\Dv_i=\diag(D_{i1},\ldots,D_{in})$, and $D_{ij}\stackrel{\rm i.i.d.}{\sim} {\rm  Bern}(p)$. Set $\mu=\frac{1}{p-p^2}$, and let ${\mvec{x}}^{t}$ denote the output  of Algorithm \ref{algo:PGD} at iteration $t$.
Then, given  $\lambda>0$, for $t=0,1,\ldots$, either ${1\over nB}\Lp{\tilde{\mvec{x}}-{\mvec{x}}^{t}}{2}{2}\leq  \delta$,  or 
\begin{align}
    \frac{1}{\sqrt{nB}}\|\xtv-\xv^{t+1}\|
\leq
\frac{2\lambda}{(p-p^2)\sqrt{nB}}\|\xtv-\xv^{t}\|+\frac{2 (p+(B-1)p^2+1)}{p-p^2}\sqrt{\delta},\label{eq:thm5-result}
\end{align}
with a probability larger than $1 - 2^{4nBr} \exp(-\frac{2n\lambda^2\delta^2}{B^2 \rho^4}) - (2^{2nBr} + 1) \exp(-\frac{2n\delta^2}{B^2 \rho^4})$

\end{theorem}

For the bound in Theorem \ref{thm:pgd} to be meaningful, the coefficient $\rho=\frac{2\lambda}{p-p^2}$ that connects the normalized  error at time $t+1$ and time $t$ needs to be smaller than $1$. Let $\rho<1$, and assume that 
\[
\lambda={\rho p(1-p)\over 2}
\]
Then, the result of Theorem \ref{thm:pgd}  implies that as $t$ grows without bound, for small enough number of frames, with high probability, the error is upper bounded by 
\begin{align}
 \lim_{t\to\infty} \frac{1}{\sqrt{nB}}\|\xtv-\xv^{t}\|
&\leq
\Big({1 \over 1- \rho }\Big)  \frac{2 (p+(B-1)p^2+1)}{p-p^2}\sqrt{\delta}.
\end{align}
It can be observed that for any choice of $\rho\in(0,1)$, the right hand side of \eqref{eq:thm5-result} is minimized at $p^*<0.5$.





\section{Experimental results}
\label{sec:experiments}

\subsection{Datasets and algorithms}

In the previous sections, we theoretically characterized the performance of compression-based SCI recovery, including both the idealized optimization and the PGD algorithm, and analyzed the effect of masks' parameters on the recovery performance. To further illustrate the impact of mask parameters on the performance of SCI recovery methods, this section focuses on video SCI and optimizes the recovery performance as a function of $p$ under the studied mask models. 

While the compression-based framework provides a robust foundation for theoretical analysis, in our simulations, we impose the source model using regularizers commonly employed in the video SCI literature instead of compression codes. Specifically, we use methods such as GAP-TV \cite{yuan2016generalized} and pretrained deep denoisers like PnP-FastDVDnet \cite{PnP_fastdvd}, as these approaches offer greater flexibility.


The algorithms are applied to simulated videos. We used six benchmark grayscale datasets: \texttt{Kobe}, \texttt{Runner}, \texttt{Drop}, \texttt{Traffic}, \texttt{Aerial}, and \texttt{Vehicle}, each with a spatial resolution of $256 \times 256$. Additionally, color datasets \texttt{ShakeNDry}, \texttt{Jockey}, \texttt{Traffic}, \texttt{Beauty}, \texttt{Runner}, and \texttt{Bosphorus} were used, each with a spatial resolution of $512 \times 512$. For each dataset, $8$ video frames (denoted as $B=8$) are compressed into a single measurement. For parameter settings, GAP-TV is run for 60 iterations, and PnP-FastDVDnet for 120 iterations per batch.

\begin{figure}[t]
\centering
\begin{subfigure}{.5\textwidth}
  \centering
  \includegraphics[width=6cm]{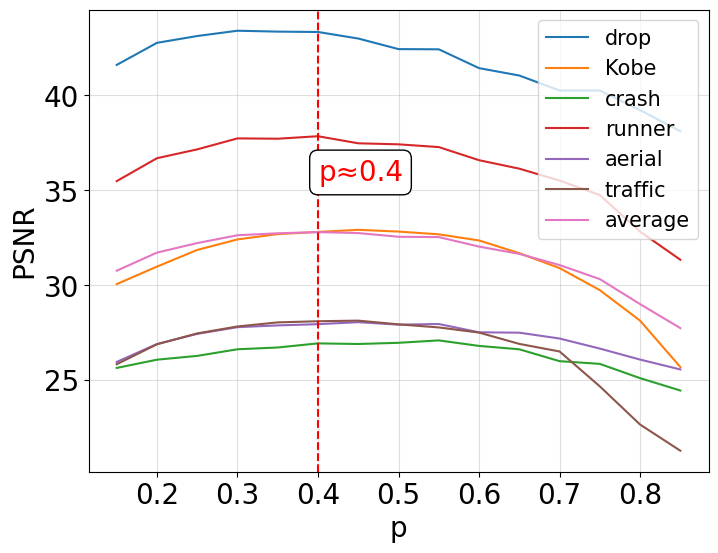}
  \caption{greyscale}
  \label{fig:sub1}
\end{subfigure}%
\begin{subfigure}{.5\textwidth}
  \centering
  \includegraphics[width=6cm]{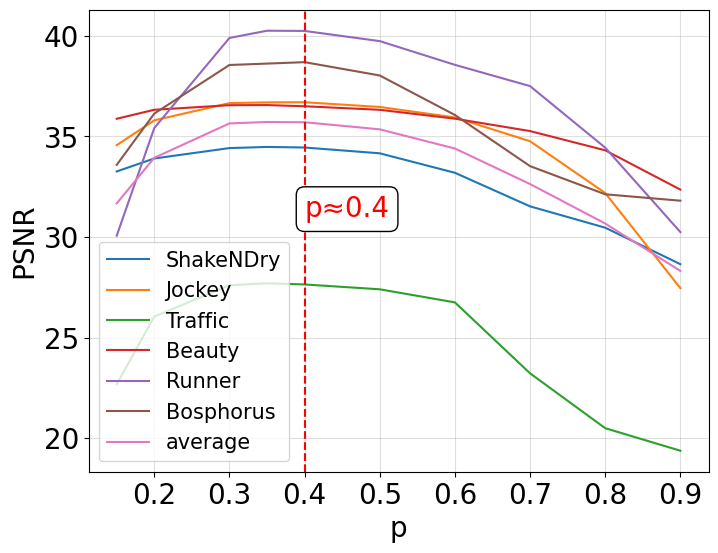}
  \caption{colorscale}
  \label{fig:sub2}
\end{subfigure}
\vspace{-0.2in}
\caption{Testing PSNR reconstructed video modulated with different masks with PnP-FastDVDnet\cite{PnP_fastdvd}}
\label{fig:4}
\end{figure}

\subsection{Performance analysis with i.i.d.~binary masks}

As the first scenario, we randomly sampled mask values $D_{ij} \overset{\mathrm{iid}}{\sim} \mathrm{Bern}(p)$, incrementally varying $p$ from 0.1 to 0.9 in steps of 0.1.

The simulation results in Fig.~\ref{fig:4}, obtained using PnP-FastDVDnet \cite{PnP_fastdvd}, align closely with the corollary of Theorem~\ref{thm:1}. It can be observed that to minimize the upper bound on the reconstruction error (and thereby optimize performance), the value of $p$ should be less than 0.5. The optimal value, $p^* \approx 0.4$, achieves the best reconstruction performance, consistent with prior studies \cite{zhang2022herosnet, iliadis2020deepbinarymask} that involve joint training of reconstruction algorithms and masks. Conversely, as $p$ approaches either 0 or 1, the PSNR drops significantly, indicating a substantial decline in recovery quality under these conditions.




\begin{figure}
\begin{center}
    \subfloat[][PSNR, shown as y-axis, of $\|\xv-\xhv\|_2$ under measuring with different frames.]{
        \includegraphics[width=15cm]{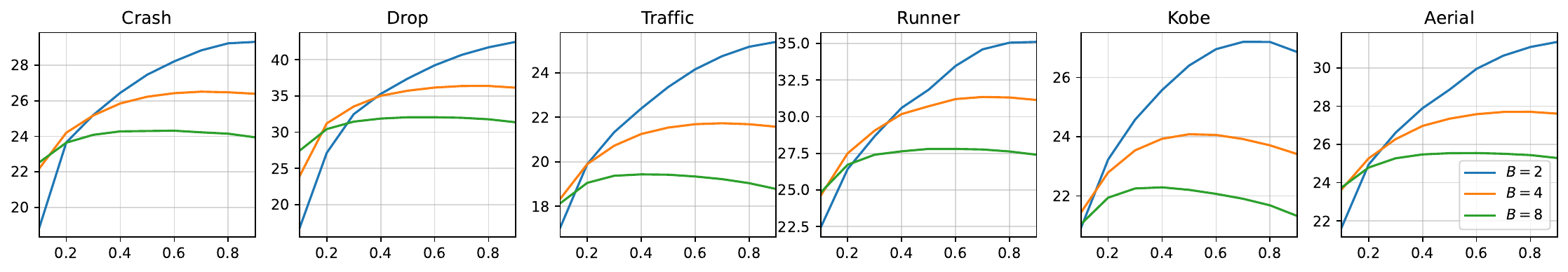}
        \label{fig:psnrgaptvframe}
    }
    \vspace{0.5cm} 
    \\
    \subfloat[][{\color{blue}Blue}, {\color{orange}orange}, and {\color{darkgreen}green} lines represent frame sizes $B=2$, $4$, and $8$, respectively. Dashed and dotted lines represent $(1-p)\|\xhv-\xv\|_2^2$ and $p\|\xbhv-\xbv\|_2^2$. GAP-TV reconstruction error terms for different frame sizes.]{
        \includegraphics[width=13cm]{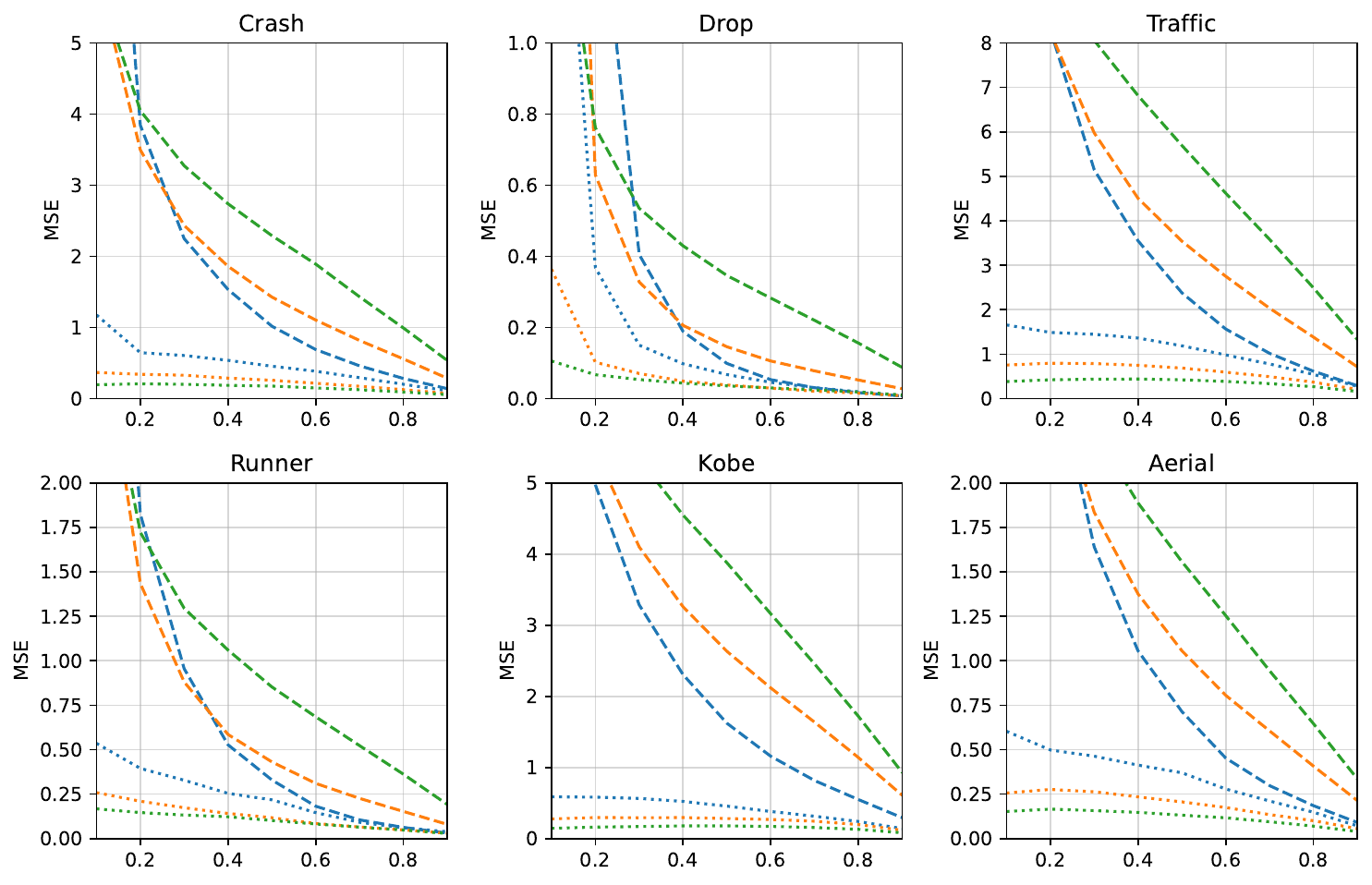}
        \label{fig:msegaptvframe}
    }
\end{center}
\vspace{-0.1in}
\caption{(a) PSNR for different frames and (b) MSE for various frame sizes.}
\label{fig:combined_figure}
\end{figure}

A much-less investigated issue in the literature is the effect of the number of frames $B$ that are mapped to a single 2D projection on the SCI performance and also its intentional interactions with the optimal mask parameters. First, intuitively, as the number of frames $B$ decreases, one expects the complexity of the problem to decrease as well, leading to a better reconstruction quality. Note that the sampling rate in an SCI system is $n/(nB)=1/B$. 
 Fig~\ref{fig:psnrgaptvframe} shows the reconstruction PSNR corresponding to different values of $B$, ranging from $B=2$ to $B=8$. It can be observed that the optimal values of $p$ are a decreasing function of $B$. That is, the higher the  number of frames, the lower the corresponding optimized values of $p$. 
 
 Furthermore, it can be observed that for very small values of $B$, such as $B=2$, the optimal values of $p$ can  exceed $0.5$. This might at first seem contradictory with the claim of Theorem \ref{thm:1}, which states that the upper bound is always minimized at some $p^*<0.5$. However, the same theorem, in fact, explains this result as well.  Recall that the error bound in Theorem \ref{thm:1} states that 
\begin{align}
(1-p)e+p\bar{e}
\leq(1+(B-1)p)\delta+\frac{\rho^2(\epsilon_1+\epsilon_2)}{p},\label{eq:sim-1}
\end{align}
where $e$ and $\bar{e}$ are defined in \eqref{eq:def-e}. The left-hand side (LHS) has two terms: $(1-p)e={1-p\over nB}\|\xhv - \xv\|_2^2$ and $p\bar{e}={p\over n}\|\xbhv - \xbv\|_2^2$. If $B$ is large enough, intuitively, one expects to be able to recover the mean of the input frames $\xbv$ more accurately, implying that $\bar{e}$ is small. To see this, note that 
\[
\E[{1\over B}\yv]=\E[{1\over B}\sum_{i=1}^B\Dv_i\xv_i]=p\xbv. 
\]
In other words, for larger values of $B$, the normalized version of $\yv$ itself is an unbiased estimator of $\xbv$. In such cases, 
the term $p\bar{e}$ becomes negligible compared to the first term, and ignoring it, as done in the corollaries of Theorem  \ref{thm:1}, has negligible impact. However, for  smaller values of $B$,  estimating $\xbv$ is more challenging, and the ultimate error is dominated by $\bar{e}$. Note that from \eqref{eq:sim-1}, since both terms are positive, we have 
\begin{align}
\bar{e}
\leq {1+(B-1)p\over p}\delta+\frac{\rho^2(\epsilon_1+\epsilon_2)}{p^2}.\label{eq:sim-2}
\end{align}
Unlike the bound in Theorem \ref{thm:1}, \eqref{eq:sim-2} is a decreasing function of $p$. These explanations are further confirmed in the simulation results that are reported in Fig~\ref{fig:msegaptvframe}.
To focus on the impact of the parameters and disentangle the potential influence of the model's capacity to represent the source structure from the effect of the SCI system parameters, we use a classical regularizer in this simulation.  When we set the number of frames $B$ at $8$ or larger, and $p$ is relatively small, the second term becomes much smaller than the first. However, for smaller values like $B=2$, as $p $ approaching to $ 1$, the difference between $(1-p)e$ and $p\bar{e}$ diminishes, and the effect of the second is no longer negligible. This MSE analysis completes our empirical study of the implications of Theorem 1 for optimizing i.i.d.~ binary-valued masks.

\subsection{Dependent mask}

\begin{figure}
\begin{center}
\includegraphics[width=8cm]{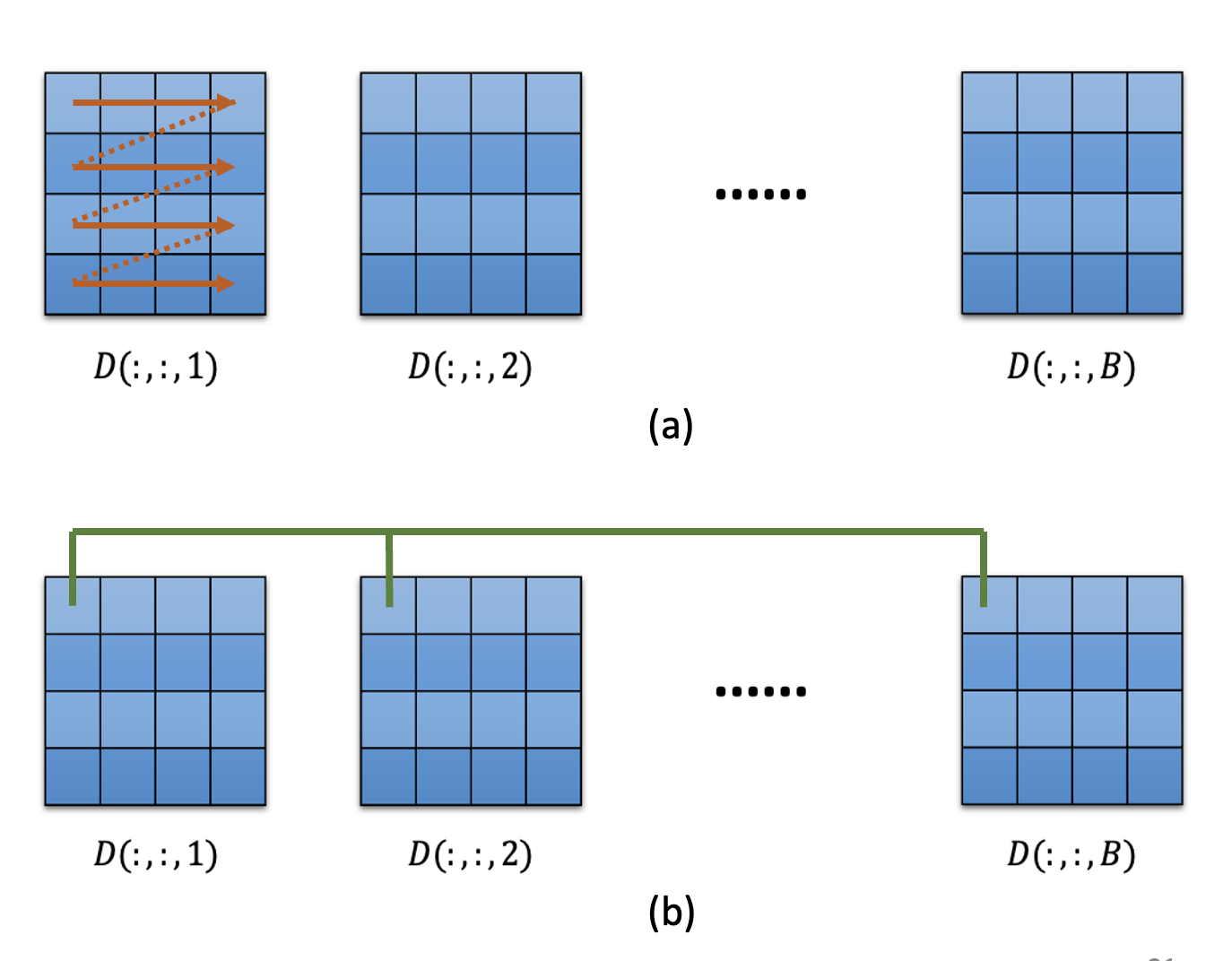}
\vspace{-0.1in}
\caption{(a) \textbf{In-frame dependence}: The mask elements corresponding to each frame form a Markov chain, and the masks of different frames are independent of one another.(b) \textbf{Out-of-frame dependence}: The mask elements corresponding to each pixel form a Markov chain, and the mask elements corresponding to different pixels are otherwise independent.}
\label{fig:5}
\end{center}
\end{figure}
In this section, we investigate the impact of dependence among the elements of the masks, focusing on both in-frame and out-of-frame dependencies, as illustrated in Fig.~\ref{fig:5} and analyzed in Theorems \ref{thm:2} and \ref{thm:3}. The dependence is modeled as a first-order stationary Markov chain with a state space of $\{0,1\}$ and transition probabilities $q_0$ and $q_1$. Based on observations from the previous section and prior work~\cite{zhang2022herosnet,iliadis2020deepbinarymask}, we have established and demonstrated that the optimized $p^*$ is approximately $0.4$. Thus, we fix the overall probability $\Pr(D_{ij}=1) = 0.4$ and vary the values of $q_0$ and $q_1$ to study their effects.

As in the previous case, we implement GAP-TV and PnP-FastDVDnet as optimization algorithms under various mask configurations. The corresponding results are reported in Table~\ref{Tab: MCdvd} and Table~\ref{Tab: MCtv}. For both types of dependence—namely in-frame and out-of-frame—we observe that increasing the transition probabilities \(q_0\) and \(q_1\) up to $0.4$ and $0.6$, respectively, leads to an overall improvement in PSNR for the benchmark datasets. Notably, when \(q_0 = 0.4\) and \(q_1 = 0.6\), the masks approximate the i.i.d. case, achieving the best performance. These simulation results consistently align with the theoretical predictions from our theorems.

\begin{table}[htbp]
\begin{center}
\tiny
\begin{tabular}{c c|c c c c c c} 
\toprule
    Type            & Transition Probability          & drop    & Kobe    & crash   & runner  & aerial  & traffic \\
\midrule 
    In-frame  & $\Pr(1|0)=0.13$~~$\Pr(0|1)=0.2$ & 41.1225 & 30.9190 & 24.3579 & 35.7955 & 26.0689 & 24.9422 \\
                    & $\Pr(1|0)=0.20$~~$\Pr(0|1)=0.3$ & 42.1855 & 31.8104 & 24.8540 & 36.3265 & 26.9808 & 26.2579 \\
                    & $\Pr(1|0)=0.25$~~$\Pr(0|1)=0.4$ & 42.6284 & 32.0601 & 25.4351 & 37.2239 & 27.4018 & 26.8268 \\
                    & $\Pr(1|0)=0.33$~~$\Pr(0|1)=0.5$ & 42.9457 & 32.5072 & 25.6517 & 37.5549 & 27.7165 & 27.3393 \\ 
\midrule
    Cross-frame   & $\Pr(1|0)=0.13$~~$\Pr(0|1)=0.2$ & 41.7627 & 29.5709 & 24.4944 & 35.1638 & 25.3162 & 25.0649 \\
                        & $\Pr(1|0)=0.20$~~$\Pr(0|1)=0.3$ & 42.7804 & 31.2580 & 25.4982 & 36.2007 & 26.5311 & 26.2431 \\
                        & $\Pr(1|0)=0.25$~~$\Pr(0|1)=0.4$ & 42.9967 & 31.9098 & 25.3333 & 36.9099 & 27.1271 & 26.8166 \\
                        & $\Pr(1|0)=0.33$~~$\Pr(0|1)=0.5$ & 43.3107 & 32.4381 & 25.5802 & 37.3619 & 27.7391 & 27.3917 \\ 
\midrule
    i.i.d.~binary       & $\Pr(1|0)=0.4$~~$\Pr(0|1)=0.6$  & 43.1761 & 32.6869 & 25.7462 & 37.7378 & 27.8686 & 27.5570 \\ 

\bottomrule
    
    \end{tabular}
\end{center}
\vspace{-0.1in}
\caption{Experiment results of different transition probability in the condition of fixing overall $\Pr(D_{ij}=1)=0.4$ with PnP-FastDVD. We report the PSNR under multiple mask assumptions.}
\label{Tab: MCdvd}
\end{table}

\begin{table}[htbp]
\begin{center}
\tiny
\begin{tabular}{c c|c c c c c c} 
\toprule
    Type                & Transition Probability          & drop    & Kobe    & crash   & runner  & aerial  & traffic \\
\midrule 
    In-frame      & $\Pr(1|0)=0.13$~~$\Pr(0|1)=0.2$ & 31.8264 & 22.2807 & 24.3003 & 27.7233 & 25.4508 & 19.4698 \\
                        & $\Pr(1|0)=0.20$~~$\Pr(0|1)=0.3$ & 31.9862 & 22.2735 & 24.2783 & 27.7797 & 25.4251 & 19.4659 \\
                        & $\Pr(1|0)=0.25$~~$\Pr(0|1)=0.4$ & 31.9638 & 22.2721 & 24.2904 & 27.6966 & 25.414  & 19.4529 \\
                        & $\Pr(1|0)=0.33$~~$\Pr(0|1)=0.5$ & 31.8759 & 22.2942 & 24.2618 & 27.7238 & 25.4567 & 19.4617 \\ 
\midrule
    Cross-frame   & $\Pr(1|0)=0.13$~~$\Pr(0|1)=0.2$ & 30.9674 & 22.2751 & 23.9964 & 27.0745 & 25.1519 & 19.3293 \\
                        & $\Pr(1|0)=0.20$~~$\Pr(0|1)=0.3$ & 31.6375 & 22.4478 & 24.2059 & 27.4941 & 25.3865 & 19.5217 \\
                        & $\Pr(1|0)=0.25$~~$\Pr(0|1)=0.4$ & 31.7824 & 22.4651 & 24.3191 & 27.6049 & 25.446  & 19.5717 \\
                        & $\Pr(1|0)=0.33$~~$\Pr(0|1)=0.5$ & 31.9233 & 22.3926 & 24.2941 & 27.6864 & 25.4701 & 19.532  \\ 
\midrule
    i.i.d.~binary       & $\Pr(1|0)=0.4$~~$\Pr(0|1)=0.6$  & 31.905  & 22.2868 & 24.2744 & 27.6372 & 25.4776 & 19.4344 \\ 

\bottomrule
    
    \end{tabular}
\end{center}
\vspace{-0.1in}
\caption{Experiment results of different transition probability in the condition of fixing overall $\Pr(D_{ij}=1)=0.4$ with GAP-TV. We report the PSNR under multiple mask assumptions.}
\label{Tab: MCtv}
\end{table}


\section{Proofs}
\label{sec:proofs}
\subsection{Proof of Theorem \ref{thm:1}} 

Let $\Tilde{\xv}=g(f(\xv))$. By assumption, the code operates at distortion $\delta$. Hence, ${1\over nB}\| \xv-\Tilde{\xv}\|_2^2\leq\delta$.  On the other hand, since $\hat{\xv}=\arg\min_{\cv\in\Cc}\|\ \yv-\sum\nolimits_{i=1}^B\Dv_i\cv_i\|_2^2$, and $\Tilde{\xv}\in \Cc$, $\|\yv-\sum_{i=1}^B\Dv_i\hat{\xv}_i\|_2\leq\|\yv-\sum_{i=1}^B\Dv_i\Tilde{\xv}_i\|_2$. But  $\yv=\sum\nolimits_{i=1}^B\Dv_i\xv_i$. Therefore, 
\begin{equation}
\|\sum_{i=1}^B\Dv_i(\xv_i-\hat{\xv}_i)\|_2\leq\|\sum_{i=1}^B\Dv_i(\xv_i-\Tilde{\xv}_i)\|_2.\label{eq:1}
\end{equation}
Note that, for a fixed $\cv\in \Cc$,
\begin{equation}
\|\sum_{i=1}^B\Dv_i(\xv_i-\hat{\xv}_i)\|_2^2=\sum_{j=1}^n(\sum_{i=1}^BD_{ij}(x_{ij}-c_{ij}))^2=\sum_{j=1}^n U_j
\end{equation}

Given a fixed $\xv$ and $\cv$, for $j=1, \ldots, n$, let 
\[
U_j=(\sum\nolimits_{i=1}^BD_{ij}(x_{ij}-c_{ij}))^2.
\]   $U_1,\ldots,U_n$ are independent  random variables and 
\begin{align}
\allowdisplaybreaks
&\E[U_j]
=\E[\sum\limits_{i=1}^B\sum\limits_{i'=1}^BD_{ij}D_{i'j}(x_{ij}-c_{ij})(x_{i'j}-c_{i'j})]\nonumber\\
&=\sum\limits_{i=1}^B\sum\limits_{i'=1 i'\neq i}^Bp^2(x_{ij}-c_{ij})(x_{i'j}-c_{i'j})+\sum\limits_{i=1}^Bp(x_{ij}-c_{ij})^2\nonumber\\
&=p^2(\sum\limits_{i=1}^B(x_{ij}-c_{ij}))^2+(p-p^2)\sum\limits_{i=1}^B(x_{ij}-c_{ij})^2. \label{eq:exp}
\end{align}

Given $\epsilon_1 > 0$ and $\epsilon_2>0$, $\xv_i\in\mathbb{R}^{n}$ and $\xv\in\mathbb{R}^{Bn}$, define events $\Ec_1$ and $\Ec_2$ as
\begin{align}
\Ec_1=\{&\frac{1}{n}\|\sum_{i=1}^B\Dv_i(\xv_i-\tilde{\xv}_i)\|_2^2
\leq\frac{p^2}{n}\|\sum_{i=1}^B(\xv_i-\tilde{\xv}_i)\|_2^2+\frac{p-p^2}{n}\|\xv-\tilde{\xv}\|_2^2+B\rho^2\epsilon_1\}\label{eq:E1}
\end{align}
and
\begin{align}
\Ec_2=\{&\frac{1}{n}\|\sum_{i=1}^B\Dv_i(\xv_i-\cv_i)\|_2^2\geq\frac{p^2}{n}\|\sum_{i=1}^B(\xv_i-\cv_i)\|_2^2+\frac{p-p^2}{n}\|\xv-\cv\|_2^2-B\rho^2\epsilon_2:\forall\cv\in \Cc\},
\label{eq:E2}
\end{align}
respectively. Define, 
\begin{align}
e={1\over nB}\|\xv-\hat{\xv}\|_2^2\;\; {\rm and}\; \bar{e}={1\over n}\|\xbv-\xbhv\|_2^2.\label{eq:def-e}
\end{align}
Then, conditioned on $\Ec_1\cap\Ec_2$, since $\hat{\xv}\in \Cc$ and $\Tilde{\xv}\in \Cc$, it follows from \ref{eq:1} that
\begin{align}
\frac{p-p^2}{n}\|\xv-\hat{\xv}\|_2^2+\frac{p^2}{n}\|B(\xbv-\xbhv)\|_2^2
&\leq\frac{p-p^2}{n}\|\xv-\Tilde{\xv}\|_2^2+\frac{p^2}{n}\|\sum_{i=1}^B(\xv_i-\tilde{\xv}_i)\|_2^2\nonumber+B\rho^2(\epsilon_1+\epsilon_2)\nonumber\\
(1-p)Be+Bp\bar{e}
&\leq\frac{1-p}{n}\|\xv-\Tilde{\xv}\|_2^2+\frac{p}{n}\|\sum_{i=1}^B(\xv_i-\tilde{\xv}_i)\|_2^2\nonumber+\frac{B\rho^2(\epsilon_1+\epsilon_2)}{p}\nonumber\\
(1-p)e+p\bar{e}
&\leq(1+(B-1)p)\delta+\frac{\rho^2(\epsilon_1+\epsilon_2)}{p},
\end{align}
where the last line follows because $\|\sum_{i=1}^B(\xv_i-\tilde{\xv}_i)\|_2^2\leq B\|\xv-\tilde{\xv}\|_2^2$ and by assumption ${1\over nB}\|\xv-\Tilde{\xv}\|_2^2\leq \delta$.


In the rest of the proof, we focus on bounding $P(\Ec_1^c\cup\Ec_2^c)$. Note that since by assumption the $\ell_{\infty}$-norm of all signals in $\Qc$ are upper-bounded by $\rho / 2$,  $U_i$'s are also bounded as 
\begin{align}
U_j
&\leq\sum_{i=1}^BD_{ij}^2\cdot\sum_{i=1}^B(x_{ij}-c_{ij})^2\nonumber\\
&\leq\sum_{i=1}^B1\cdot\sum_{i=1}^B(\frac{\rho}{2}+\frac{\rho}{2})^2=B^2\rho^2.\label{eq:bound-Uj}
\end{align}


Therefore, applying the Hoeffding's inequality, 
\begin{align}
&\Pr(\frac{1}{n}\sum_{j=1}^nU_j\geq\frac{1}{n}\E[\sum_{j=1}^nU_j]+B\rho^2\epsilon_1)\leq\exp(-\frac{2n^2B^2\rho^4\epsilon_1^2}{n(B^2\rho^2)^2})=\exp(-\frac{2n\epsilon_1^2}{B^2}). 
\label{eq:C1}
\end{align}
similarly,
\begin{align}
&\Pr(\frac{1}{n}\sum_{i=j}^nU_j\leq\frac{1}{n}\E[\sum_{j=1}^nU_j]-B\rho^2\epsilon_2)\leq\exp(-\frac{2n^2B^2\rho^4\epsilon_2^2}{n(B^2\rho^2)^2})=\exp(-\frac{2n\epsilon_2^2}{B^2}).
\label{eq:C2}
\end{align}

Therefore, 
\begin{equation}
\Pr(\Ec_1^c)
\leq\exp(-\frac{2n\epsilon_1^2}{B^2})\label{eq:E1c-prob}
\end{equation}
and, by the union bound, since  $|\Cc|\leq 2^{Br}$,
\begin{equation}
\Pr(\Ec_2^c)
\leq 2^{Br}\exp(-\frac{2n\epsilon_2^2}{B^2}).\label{eq:E2c-prob}
\end{equation}
Again by the union bound, $P(\Ec_1\cap\Ec_2)\geq 1-P(\Ec_1^c)-P(\Ec_2^c)$. Setting $\epsilon_1=\epsilon_2=\epsilon/2$,
\begin{align}
(p-p^2)\|\xv-\hat{\xv}\|_2^2&\leq(p+(B-1)p^2)\|\xv-\Tilde{\xv}\|_2^2+nB\rho^2\epsilon_1+nB\rho^2\epsilon_2\nonumber\\
&\leq(p+(B-1)p^2)nB\delta+nB^2\rho^2\epsilon.
\end{align}
Also, from \ref{eq:E1c-prob} and \ref{eq:E2c-prob},
\begin{align*}
P(\Ec_1\cap\Ec_2)&\geq 1-\exp(-\frac{2n\epsilon_1^2}{B^2})-2^{Br}\exp(-\frac{2n\epsilon_2^2}{B^2})\\
&=1-(2^{Br}+1)\exp(-\frac{2nB^2\epsilon^2}{4B^2})\\
&=1-(2^{Br}+1)\exp(-n\epsilon^2/2).
\end{align*}
Setting 
\[
\epsilon=\sqrt{2\ln(2) Br+2\eta r\over n},
\]
we have
\begin{align*}
\ln ((2^{Br}+1)\exp(-n\epsilon^2/2)) 
&\leq  Br \ln 2+\ln 2 -(\ln 2 Br+\eta r)=\ln 2-\eta r,
\end{align*}
which implies that 
\[
P(\Ec_1\cap \Ec_2)\geq 1-2\ex^{-\eta r}.
\]
Also, from \eqref{eq:main-thm1-result}, noting $2\ln 2\leq 2$,
\begin{align}
{1\over nB}\|\xv-\hat{\xv}\|_2^2
&\leq\Big(1+ {Bp \over 1-p}\Big) \delta+ {B\rho^2\sqrt{2(B+\eta)}\over p-p^2}\sqrt{ r\over n},
\end{align}
which is the desired result.

To optimize the upper bound with respect to $p$, define
\[
u(p)=\Big(1+ {Bp \over 1-p}\Big) \delta+ {a\over p-p^2}
\] 
where $a=B\rho^2\sqrt{2r(B+\eta)/n}$. Note that $u(0)=u(1)=\infty$. Let $p^*$ denote the value of $p\in(0,1)$ that minimizes $u(p)$. Note that 
\begin{align}
u'(p)&={B\delta\over (1-p)^2}+a({1\over (1-p)^2}-{1\over p^2}).
\label{eq:thm1pfunc}
\end{align}
Therefore, $\lim_{p\to 0} u'(p)= -\infty$ and  $u'({1\over 2})=4B\delta>0$, which implies that $p^*$ where $u'(p^*)=0$ belongs to $(0,{1\over 2})$.  
\[
p^*={1\over 1+\sqrt{1+{\delta\over \rho^2}\sqrt{n\over 2r(B+\eta)}}}
\]

\subsection{Proof of Corollary \ref{cor:1}}
The proof follows similar to the proof of Theorem \ref{thm:1}. The only difference is that, here, 
\begin{align}
    \E[U_j]
&=\E[\sum\limits_{i=1}^B\sum\limits_{i'=1}^BD_{ij}D_{i'j}(x_{ij}-c_{ij})(x_{i'j}-c_{i'j})]\nonumber\\
&=(2p-1)^2(\sum\limits_{i=1}^B(x_{ij}-c_{ij}))^2+4(p-p^2)\sum\limits_{i=1}^B(x_{ij}-c_{ij})^2.
\end{align}
The defined events can be updated using this new expected value. But the concentration bounds remain in tact, as the uppper bound on $U_j$s remains the same.

\subsection{Proof of Corollary \ref{cor:1.2}}
The proof follows similar to the proof of Theorem \ref{thm:1}. The only difference is that, here, \begin{align}
    \E[U_j]
&=\E[\sum\limits_{i=1}^B\sum\limits_{i'=1}^BD_{ij}D_{i'j}(x_{ij}-c_{ij})(x_{i'j}-c_{i'j})]\nonumber\\
&=p^2(\sum\limits_{i=1}^B(x_{ij}-c_{ij}))^2+(q-p^2)\sum\limits_{i=1}^B(x_{ij}-c_{ij})^2.
\end{align}
The variables remain bounded as before and the same concentration bounds can be applied here too. 



\subsection{Proof of Theorem \ref{thm:2}}
Following the same steps as  the initial steps of proof of Theorem \ref{thm:1}, we have
\[
\sum_{j=1}^n(\sum_{i=1}^BD_{ij}(x_{ji}-\hat{x}_{ij}))^2\leq\sum_{j=1}^n(\sum_{i=1}^BD_{ij}(x_{ij}-\tilde{x}_{ij}))^2
\]
Define $\dv_j=[D_{1j},\ldots,D_{Bj}].$
Note that $\dv_1,\ldots,\dv_n$ is a stationary first Markov process with state space $\Sc=\{0,1\}^B$ such that 
\[
p(\dv_i|\dv_1,\ldots,\dv_{i-1}) =p(\dv_i|\dv_{i-1}) =\prod_{j=1}^Bp(d_{ij}|d_{(i-1)j}),
\]
where $p(d_{ij}|d_{(i-1)j})$ agrees with the transition probability of the Markov chain used for generating the masks. 
Given $\xv$ and $\cv$,  for $j=1, \ldots, n$, let $\varphi(\dv_j)=(\sum\nolimits_{i=1}^BD_{ij}(x_{ij}-c_{ij}))^2$.  Unlike in the proof of Theorem \ref{thm:1}, $\varphi(\dv_1),\ldots,\varphi(\dv_n)$ are no longer independent. However, the expected values of $\varphi(\dv_j)$'s are the same as those of $U_j$'s, because the dependencies are in-frame. Therefore, 
\[
\E[\varphi(\dv_j)]
=p^2(\sum\limits_{i=1}^B(x_{ij}-c_{ij}))^2+(p-p^2)\sum\limits_{i=1}^B(x_{ij}-c_{ij})^2. \nonumber
\]
Similar to the proof of Theorem \ref{thm:1}, define events $\Ec_1$ and $\Ec_2$, as \eqref{eq:E1} and \eqref{eq:E2}, respectively. 
To bound $\Pr((\Ec_1\cap\Ec_2)^c)$, we need  to show the concentration of $\sum_{j=1}^n\varphi(\dv_j)$ around its expected value. To achieve this goal, we use a result from \cite{kontorovich2008concentration}, which is explained in Appendix A of the extended version of the paper \cite{zhao2023theoretical}. To employ Theorem 5 in Appendix A , note that since the Markov chain is assumed to be stationary, $\theta_1=\theta_2=\ldots =\theta_{n-1}$. Therefore,
\begin{align}
    M_n &= \max_{1\leq i\leq n-1}(1+\theta_i+\theta_i\theta_{i+1}+\cdots+\theta_i\cdots\theta_{n-1})\nonumber\\
    &=1+\theta_1+\theta_1^2+\theta_1^3+\cdots+\theta_1^{n-1}\nonumber\\
    &=\frac{1-\theta_1^n}{1-\theta_1}.
\end{align}




To use  theorem stated in Appendix A of \cite{zhao2023theoretical}, let $c$ denote the Lipschitz coefficient  of function $\varphi:\Sc\to\mathbb{R}$, defined earlier. Then, we have
\begin{align}
&\Pr\big(\;\frac{1}{n}\sum_{j=1}^n\varphi(D_j)\geq\frac{1}{n}\E\left[\varphi(D_j)\right]+B\rho^2\epsilon_1\big)\nonumber\\
&\leq\exp(-\frac{n^2B^2\rho^4\epsilon_1^2}{2nc^2M_n^2})\nonumber\\
&\leq\exp(-\frac{nB^2\rho^4\epsilon_1^2}{2c^2}(1-\theta_1)^2),\label{eq:bd-E1}
\end{align}
and
\begin{align}
&\Pr\big(\;\frac{1}{n}\sum_{j=1}^n\varphi(D_j)\leq\frac{1}{n}\E\left[\varphi(D_j)\right]-B\rho^2\epsilon_2\big)\nonumber\\
&\leq\exp(-\frac{n^2B^2\rho^4\epsilon_2^2}{2nc^2M_n^2})\nonumber\\
&\leq\exp(-\frac{nB^2\rho^4\epsilon_2^2}{2c^2}(1-\theta_1)^2),\label{eq:bd-E2}
\end{align}
where in deriving both bounds we have used the fact that  $M_n={1-\theta_1^n \over 1-\theta_1}\leq {1\over 1-\theta_1}.$
To bound the Lipschitz constant $c$, note that for and $\dv_j,\dv'_j\in\{0,1\}^B$, we have 
\begin{align}
    &|\varphi(\dv_j)-\varphi(\dv'_j)|\nonumber\\
    &=|(\sum_{i=1}^BD_{ij}(x_{ij}-c_{ij}))^2-(\sum_{i=1}^BD_{ij}'(x_{ij}-c_{ij}))^2|\nonumber\\
    &=|\sum_{i=1}^B (D_{ij} + D_{ij}')(x_{ij}-c_{ij})| \cdot |\sum_{i=1}^B (D_{ij} - D_{ij}')(x_{ij}-c_{ij})|\nonumber\\
    &\stackrel{\rm (a)}{\leq} 2B\rho^2 d_H(\dv_j,\dv'_j),
\end{align}
where (a) follows because for all $\xv\in\Qc$, $\|\xv\|_{\infty}\leq {\rho \over 2}$. This implies that $c\geq 2B\rho^2$. 
Finally, setting $\epsilon_1=\epsilon_2=\epsilon/2$, and noting that  $| \Cc |\leq 2^{Br}$ yields the desired result. 

%
%
%



\subsection{Proof of Theorem \ref{thm:3}}
Again we follow the same steps as  the initial steps of proof of Theorem \ref{thm:1} to derive $\sum_{j=1}^n(\sum_{i=1}^BD_{ij}(x_{ji}-\hat{x}_{ij}))^2\leq\sum_{j=1}^n(\sum_{i=1}^BD_{ij}(x_{ij}-\tilde{x}_{ij}))^2.$
As in the proof of Theorem \ref{thm:2},  define $\dv_j=[D_{1j},\ldots,D_{Bj}].$ Unlike the proof of Theorem \ref{thm:2}, here $\dv_1,\ldots,\dv_n$ are independent and identically distributed. Again similar to the proof of Theorem \ref{thm:1}, given $\xv$ and $\cv$, for $j=1, \ldots, n$,  define 
\[
U_j(\xv,\cv)=(\sum\nolimits_{i=1}^BD_{ij}(x_{ij}-c_{ij}))^2.
\]  Note that   $U_1(\xv,\cv),\ldots,U_n(\xv,\cv)$ are independent  random variables. Moreover, they are positive and  bounded with the same upper bound as the one derived in \eqref{eq:bound-Uj}. Therefore, we can still apply the Hoeffding's inequality and derive \eqref{eq:C1} and \eqref{eq:C2}. The key difference now is that computing $\E[U_j(\xv,\cv)]$ is more complex as the entries of $\dv_j$ are not independent.  

To compute $\E[U_j(\xv,\cv)]$, define $\mu_{ij}=x_{ij}-c_{ij}$. Also, note that as before, $\E[D_{ij}^2]=\E[D_{ij}]=p$. Moreover,
\begin{align}
\E[U_j(\xv,\cv)]&=\E[(\sum_{i=1}^BD_{ij}(x_{ij}-c_{ij}))^2]\nonumber\\
&=\sum_{i_1=1}^B\sum_{i_2=1}^B\E[D_{i_1j}D_{i_2j}]\mu_{i_1j}\mu_{i_2j}.\label{eq:Uj-step1}
\end{align}
Without loss of generality, assume that $i_1<i_2$. Then, $\E[D_{i_1j}D_{i_2j}]=\Pr(D_{i_1j}=D_{i_2j}=1)=\Pr(D_{i_1j}=1)\Pr(D_{i_2j}=1|D_{i_1j}=1)$. To compute $\Pr(D_{i_2j}=1|D_{i_1j}=1)$, we need to compute $(i_2-i_1)$-th order transition probability of the Markov chain. 
%
The transition kernel of the Markov chain can be written as 
\[P=\begin{bmatrix}
    1-q_0 & q_0\\
    q_1 & 1-q_1\\
\end{bmatrix}.
\]
Let $Q=\begin{bmatrix}
    1 & -q_0\\
    1 & q_1\\
\end{bmatrix}$, and let $\alpha= 1-q_0-q_1.$
Then,
\begin{gather}
    \Pi=Q\begin{bmatrix}
        1 & 0\\
        0 & \alpha\\
    \end{bmatrix}Q^{-1}
\end{gather}
Using this representation, the $k$-th order transition probability of this Markov chain can be written as 
\begin{gather}
     \Pi^k=\frac{1}{q_0+q_1}
    \begin{bmatrix}
        q_1 & q_0\\
        q_1 & q_0\\
    \end{bmatrix}
    +
    \frac{\alpha^k}{q_0+q_1}
    \begin{bmatrix}
        q_0 & -q_0\\
        -q_1 & q_1\\
    \end{bmatrix}.
\end{gather}
Therefore, for $k=1,\ldots,n-i$
\[
\Pr(D_{(i+k)j}=1|D_{ij}=1)={q_0+\alpha^k q_1\over q_0+q_1}=p+(1-p)\alpha^k. 
\]
Thus,
\begin{align}
\E&[U_j(\xv,\cv)]=\sum_{i_1}^B\sum_{i_2}^B\E[D_{i_1j}D_{i_2j}]\mu_{i_1j}\mu_{i_2j}\nonumber\\
&=\sum_{i_1}^B\sum_{i_2}^Bp(p+(1-p)\alpha^{|i_1-i_2|})\mu_{i_1j}\mu_{i_2j}\nonumber\\
&=p^2(\sum_{i}^B\mu_{ij})^2+p(1-p)\sum_{i_1}^B\sum_{i_2}^B\alpha^{|i_1-i_2|}\mu_{i_1j}\mu_{i_2j}\nonumber\\
&=p^2(\sum_{i}^B\mu_{ij})^2+p(1-p)\muv_j^T \Lambda \muv_j,
\label{eq:E-Uj-exact}
\end{align}
where $\muv_j=[\mu_{1j},\ldots,\mu_{Bj}]^T$ and $\Lambda$ is defined in \eqref{eq:def-Lambda}. Therefore, $\E[U_j(\xv,\cv)]$ can be upper- and lower-bounded as
\begin{align}
\E[U_j(\xv,\cv)]&\leq p^2(\sum_{i=1}^B\mu_{ij})^2+p(1-p)\lambda_{\max}(\Lambda) \|\muv_j\|_2^2,\label{eq:Uj-lb}
\end{align}
and
\begin{align}
\E[U_j(\xv,\cv)]&\geq p^2(\sum_{i=1}^B\mu_{ij})^2+p(1-p)\lambda_{\min}(\Lambda) \|\muv_j\|_2^2.\label{eq:Uj-ub}
\end{align}
Define,
\[
 \Ec_1=\{\sum_{j=1}^n U_j(\xv,\cv)\geq \sum_{j=1}^n \E[U_j(\xv,\cv)]-B\rho^2\epsilon_1,\;\forall \cv\in\Cc\},
\]
and
\[
 \Ec_2=\{\sum_{j=1}^n U_j(\xv,\tilde{\xv})\leq \sum_{j=1}^n \E[U_j(\xv,\tilde{\xv})]+B\rho^2\epsilon_2,\},
\]
respectively. As explained earlier, we the bounds in  \eqref{eq:C1} and \eqref{eq:C2} still hold here too. Therefore, the lower bound on $\Ec_1\cap\Ec_2$ is the same as before. But conditioned on $\Ec_1\cap\Ec_2$, employing the bounds in \eqref{eq:Uj-lb} and \eqref{eq:Uj-ub}, it follows that  
\begin{align*}
&\frac{p(1-p)\lambda_{\min}(\Lambda)}{n}\|\xv-\hat{\xv}\|_2^2\nonumber\\
&\leq\frac{p(1-p)\lambda_{\max}(\Lambda)}{n}\|\xv-\Tilde{\xv}\|_2^2+\frac{p^2}{n}\|\sum_{i=1}^B(\xv_i-\tilde{\xv}_i)\|_2^2\nonumber\\
&\;\;\;\;+B\rho^2\epsilon_1+B\rho^2\epsilon_2\nonumber\\
&\leq\frac{p(1-p)\lambda_{\max}(\Lambda)+p^2B}{n}\delta+B\rho^2\epsilon,
\end{align*}
where the last line follows by setting $\epsilon_1=\epsilon_2={\epsilon\over 2}$. 

\subsection{Proof of Corollary \ref{cor:thm-3}}
According to the Gershgorin circle theorem \cite{gershgorin1931uber}, since all the diagonal entries of $\Lambda$ are equal to one, every eigenvalue of $\Lambda$ lies within at least one of the Gershgorin discs. These discs are all centered at one and have radii equal to 
\[
r_i=\sum_{j\neq i}\Lambda_{ij},
\]
for $i=1,\ldots,B$. Therefore,
\[
1- \max_i r_i \leq \lambda_{\min}(\Lambda)\leq \lambda_{\max}(\Lambda)\leq 1+\max r_i.
\] 
But 
\[
\max r_i\leq 2(\alpha+\alpha^2+\ldots)={2\alpha\over 1-\alpha}.
\]
Therefore,
\[
{1-3\alpha\over 1-\alpha} \leq \lambda_{\min}(\Lambda)\leq \lambda_{\max}(\Lambda)\leq {1+\alpha\over 1-\alpha}.
\]
Inserting these bounds in the result of Theorem \ref{thm:3} yields the desired result.

\subsection{Proof of Theorem  \ref{thm:pgd}}
\label{sec:proof-pgd}

Define error vector and its normalized version as 
\[
\thetav^t=\xtv-\xv^t~~\text{and,}~~ \bthev^t=\frac{\ev^t}{\|\ev^t\|_2},
\]
respectively. 
For $i=1,\ldots,B$, the $i^{th}$ frame  of each of these error vectors can be written as $\thetav^t_i=\xtv_i-\xv^t_i$
\[
\bthev_i^t=\frac{1}{\|\ev^t\|_2}(\xtv_i-\xv^t_i).
\]
For all $i=1,\ldots, B$ and $j=1,\ldots,n$,  we have
\[
|\bthe^t_{ij}|^2=\frac{(\tilde{x}^t_{ij}-x^t_{ij})^2}{\|\thetav^t\|^2_2}\leq\frac{\rho^2}{nB\delta},~~\text{and}~~|\bthe^{t+1}_{ij}|^2\leq\frac{\rho^2}{nB\delta}
\]
because 1) by assumption $\min\{\|\ev^t\|^2_2,\|\ev^{t+1}\|^2_2\}\geq nB\delta$ and 2) $\|\xv\|_{\infty}\leq {\rho/2}$, for all $\xv\in\Qc$.

On the other hand, since $\xv^{t+1}$ is the closest  codeword to $\sv^{t+1}$ in $\Cc$, and $\xtv$ is also in $\Cc$, it follows
\[
\sum_{i=1}^B\|\sv_i^{t+1}-\xv_i^{t+1}\|^2_2\leq \sum_{i=1}^B\|\sv_i^{t+1}-\xtv_i\|^2_2
\]
But $\|\sv_i^{t+1}-\xv_i^{t+1}\|^2_2=\|\sv_i^{t+1}-\xtv_i+\xtv_i-\xv_i^{t+1}\|^2_2=\|\sv_i^{t+1}-\xtv_i\|^2_2+2\langle\sv_i^{t+1}-\xtv_i,\xtv_i-\xv_i^{t+1}\rangle+\|\xtv-\xv_i^{t+1}\|\leq \|\sv_i^{t+1}-\xtv_i\|^2_2$. Let 
\[
\ev^t=y-\sum_{i=1}^B \Dv_i\xv_i^t, \text{and}\quad \sv^{t+1}=\xv_i^t+\mu\Dv_i \ev^t. 
\]
It follows that 
\begin{align}
&\sum_{i=1}^B\|\xtv_i-\xv_i^{t+1}\|^2_2\leq 2\sum_{i=1}^B\langle\xtv_i-\sv_i^{t+1},\xtv_i-\xv_i^{t+1}\rangle\nonumber\\
&=2\sum_{i=1}^B\langle\xtv_i-\xv_i^t-\mu\Dv_i \ev^t,\xtv_i-\xv_i^{t+1}\rangle\nonumber\\
&=2\sum_{i=1}^B\langle\xtv_i-\xv_i^{t},\xtv_i-\xv_i^{t+1}\rangle-2\langle\sum_{i=1}^B\Dv_i(\xtv_i-\xv_i^{t}),\sum_{i=1}^B\mu\Dv_i(\xtv_i-\xv_i^{t+1})\rangle\nonumber\\
&~~~~-2\langle\sum_{i=1}^B\Dv_i(\xv_i-\xtv_i),\sum_{i=1}^B\mu\Dv_i(\xtv_i-\xv_i^{t+1})\rangle
\label{eq:pdg1}
\end{align}
We can further simplify $2\sum_{i=1}^B\langle\xtv_i-\xv_i^{t},\xtv_i-\xv_i^{t+1}\rangle-2\langle\sum_{i=1}^B\Dv_i(\xtv_i-\xv_i^{t}),\sum_{i=1}^B\mu\Dv_i(\xtv_i-\xv_i^{t+1})\rangle$ as 
\begin{align}
&2\sum_{i=1}^B\langle\xtv_i-\xv_i^{t},\xtv_i-\xv_i^{t+1}\rangle-2\mu \langle\sum_{i=1}^B\Dv_i(\xtv_i-\xv_i^{t}),\sum_{i=1}^B\Dv_i(\xtv_i-\xv_i^{t+1})\rangle\nonumber\\
&=2\|\thetav^t\|_2\|\thetav^{t+1}\|_2(\sum_{i=1}^B\langle\bthev^t_i,\bthev^{t+1}_i\rangle-\mu\langle\sum_{i=1}^B\Dv_i\bthev^t_i,\sum_{i=1}^B\Dv_i\bthev^{t+1}_i\rangle).\label{eq:thm51-bound1}
\end{align}
For $j=1,\ldots,n$, define random variables are  $U_j$ and $V+j$ as
\[
U_j= \sum_{i_1=1}^B D_{i_1 j}\bthe^t_{i_1j}\text{and}\quad V_j=\sum_{i_2=1}^B D_{i_2 j}\bthe^{t+1}_{i_2 j}.
\]
Then, using these definitions, the bound in \eqref{eq:thm51-bound1} can further be simplified as 
\begin{align}
&\sum_{i=1}^B\langle\bthev^t_i,\bthev^{t+1}_i\rangle-\mu\langle\sum_{i=1}^B\Dv_i\bthev^t_i,\sum_{i=1}^B\Dv_i\bthev^{t+1}_i\rangle\\\nonumber
&= \sum_{j=1}^n(\sum_{i=1}^B\bthe^t_{ij}\bthe^{t+1}_{ij}- \mu(\sum_{i_1=1}^B  D_{i_1 j}\bthe^t_{i_1j})(\sum_{i_2=1}^B  D_{i_2 j}\bthe^{t+1}_{i_2 j}))\nonumber\\
& =  \sum_{j=1}^n(\sum_{i=1}^B  \bthe^t_{ij}\bthe^{t+1}_{ij}- \mu U_j V_j). \label{eq:thm51-bound2}
\end{align}
Note that 
\begin{align}
&\E[U_j V_j]=(p-p^2)\sum_{i=1}^B\bthe_{ij}^t\bthe_{ij}^{t+1}+p^2(\sum_{i=1}^B\bthe_{ij}^t)(\sum_{i=1}^B\bthe_{ij}^{t+1}).
\end{align} 
Moreover, as we argued earlier, $|\bthe_{ij}^t|^2$s are bounded by $\frac{\rho^2}{nB\delta}$. Therefore, $U_jV_j$ is also bounded as,
\begin{align}
U_jV_j=\sum_{i_1=1}^B D_{i_1 j}\bthe^t_{i_1j}\sum_{i_2=1}^B D_{i_2 j}\bthe^{t+1}_{i_2 j}
\leq \sqrt{(\sum_{i_1=1}^B D_{i_1 j}^2)(\sum_{i_1=1}^B (\bthe^t_{i_1j})^2)(\sum_{i_2=1}^B D_{i_2 j}^2)(\sum_{i_2=1}^B (\bthe^{t+1}_{i_2 j})^2)}=\frac{B\rho^2}{n\delta}.
\end{align}

Adding $\sum_{j=1}^n\frac{p}{1-p} ( \sum_{i=1}^B \bthe_{ij}^t )( \sum_{i=1}^B \bthe_{ij}^{t+1} ) >0$ to the right hand side of \eqref{eq:thm51-bound2}, and noting that $\mu=1/(p-p^2)$, it follows that 
\begin{align}
&\sum_{i=1}^B\langle\bthev^t_i,\bthev^{t+1}_i\rangle-\mu\langle\sum_{i=1}^B\Dv_i\bthev^t_i,\sum_{i=1}^B\Dv_i\bthev^{t+1}_i\rangle\nonumber\\
&\leq \frac{1}{p-p^2} \sum_{j=1}^n \left( (p-p^2) \left( \sum_{i=1}^B \bthe^t_{ij} \bthe^{t+1}_{ij} + \frac{p}{1-p} \left( \sum_{i=1}^B \bthe_{ij}^t \right) \left( \sum_{i=1}^B \bthe_{ij}^{t+1} \right) \right) - U_j V_j \right)\nonumber\\
&\leq \frac{1}{p-p^2} \sum_{j=1}^n (\E[U_jV_j] - U_j V_j).\label{eq:thm51-bound3}
\end{align}


Define the set of possible normalized error vectors of interest as 
\begin{align}
\Fc \triangleq \{\frac{\cv-\cv'}{\|\cv-\cv'\|_2}~;~(\cv-\cv')\in\Cc^2~,~\|\cv-\cv'\|_2\geq \sqrt{nB\delta}\}.  
\end{align}
For  $\lambda\in(0,0.5)$, define event $\Ec_1$ as
\begin{align}
\Ec_1=\big\{ \sum_{j=1}^n \E[U_jV_j] - U_j V_j\leq \lambda ~,~\forall(\thetav,\thetav')\in\Fc\big\}.
\end{align}

To bound the second term on the right hand side of ~\eqref{eq:pdg1}, we employ the Cauchy-Schwartz inequality as follows
\begin{align}
&2\langle\sum_{i=1}^B\Dv_i(\xv_i-\xtv_i),\mu\sum_{i=1}^B\Dv_i(\xtv_i-\xv_i^{t+1})\rangle\nonumber\\
&\leq 2\mu\|\sum_{i=1}^B\Dv_i(\xv_i-\xtv_i)\|_2\cdot\|\sum_{i=1}^B\Dv_i(\xtv_i-\xv_i^{t+1})\|_2\nonumber\\
&=2\mu\|\sum_{i=1}^B\Dv_i(\xv_i-\xtv_i)\|_2\cdot\|\sum_{i=1}^B\Dv_i\thetav_i^{t+1}\|_2.\label{eq:thm51-bound4}
\end{align}
Finally, combining \eqref{eq:thm51-bound2} and \eqref{eq:thm51-bound4}, with \eqref{eq:pdg1}, we have that conditioned on $\Ec_1$, 
\begin{align}
\|\thetav^{t+1}\|_2 
\leq 2\lambda\mu \|\thetav^t\|_2 +\frac{2}{p-p^2}\|\sum_{i=1}^B\Dv_i(\xv_i-\xtv_i)\|_2\cdot\|\sum_{i=1}^B\Dv_i\bthev_i^{t+1}\|_2.\label{eq:term1}
\end{align}
For  $\epsilon_1,\epsilon_2>0$, define event $\Ec_2$ and $\Ec_3$ as
\begin{align}
\Ec_2=\left\{\|\sum_{i=1}^B\Dv_i(\xv_i-\xtv_i)\|_2^2\leq \E[\|\sum_{i=1}^B\Dv_i(\xv_i-\xtv_i)\|_2^2]+nB\rho^2\epsilon_1 \right\},
\label{eq:pgd_event2}
\end{align}
and 
\begin{align}
\Ec_3=\left\{\|\sum_{i=1}^B\Dv_i\bthev_i)\|_2^2\leq \E[\|\sum_{i=1}^B\Dv_i\bthev_i^{t+1}\|_2^2]+\epsilon_2~,~\forall\thetav\in\Fc \right\},
\label{eq:pgd_event3}
\end{align}
respectively.  Since by definition $\|\xv-\xtv\|_2^2\leq nB\delta$, conditioned on $\Ec_2$,
\begin{align}
\|\sum_{i=1}^B\Dv_i(\xv_i-\xtv_i)\|_2^2&\leq \E[|\sum_{i=1}^B\Dv_i(\xv_i-\xtv_i)\|_2^2]+nB\rho^2\epsilon_1\nonumber\\
&\leq nB((p+(B-1)p^2)\delta+\rho^2\epsilon_1).\label{eq:term2}
\end{align}
Also, conditioned on $\Ec_3$, we have
\begin{align}
\|\sum_{i=1}^B\Dv_i\bthev_i)\|_2^2\leq p+(B-1)p^2+\epsilon_2.\label{eq:term3}
\end{align}
Finally, conditioned on $\Ec_1\cap\Ec_2\cap\Ec_3$, combining the bounds in \eqref{eq:term2} and \eqref{eq:term3} with \eqref{eq:term1}, we have
\begin{align}
\|\thetav^{t+1}\|_2 
&\leq 2\lambda\|\thetav^t\|_2 +2\mu\|\sum_{i=1}^B\Dv_i(\xv_i-\xtv_i)\|_2\cdot\|\sum_{i=1}^B\Dv_i\bthev_i^{t+1}\|_2\nonumber\\
&\leq {2 \lambda\over p(1-p)}  \|\thetav^{t}\|_2+2\frac{1}{p-p^2}\sqrt{nB((p+(B-1)p^2)\delta+\rho^2\epsilon_1)(p+(B-1)p^2+\epsilon_2)},
\end{align}
or, using the definitions, or $\thetav^{t}$ and $\thetav^{t+1}$ and setting $\epsilon_1=\frac{\delta}{\rho^2}$ and $\epsilon_2=1$ we have
\begin{align}
\frac{1}{\sqrt{nB}}\|\xtv-\xv^{t+1}\|
\leq
\frac{2\lambda}{(p-p^2)\sqrt{nB}}\|\xtv-\xv^{t}\|+\frac{2 (p+(B-1)p^2+1)}{p-p^2}\sqrt{\delta},
\end{align}
which is the desired result. 


To finish the proof, we need to bound $\Pr(\Ec_1\cap\Ec_2\cap\Ec_3)$. First, employing the Hoeffding's inequality, for a fixed $(\thetav,\thetav')\in\Fc$, we have
\begin{align}
\Pr\Big( \sum_{j=1}^n \E[U_jV_j] - U_j V_j \geq \epsilon\Big)
\leq\exp(-\frac{2n\epsilon^2\delta^2}{B^2\rho^4}).\label{eq:pgd_event1}
\end{align}
Combining \eqref{eq:pgd_event1} with union bound, we have
\begin{align}
\Pr(\Ec_1^c)\leq|\Fc|^2\exp(-\frac{2\lambda^2\delta^2n}{\mu^2B^2\rho^4})\leq 2^{4nBr}\exp(-\frac{2n\lambda^2\delta^2}{B^2\rho^4})
\end{align}

\noindent Similarly, again by the Hoeffding's inequality, and since $\epsilon_1=\frac{\delta}{\rho^2}$, it follows that
\begin{align}
\Pr(\Ec_2^c)\leq\exp(-\frac{2n\epsilon_1^2}{B^2})=\exp(-\frac{2n\delta^2}{B^2\rho^4}).
\end{align}
Finally, since $(\sum_{i=1}^B\Dv_i\bthev_i^{t+1})^2$ is bounded by $\frac{B\rho^2}{n\delta}$ and $\epsilon_2=1$, we have
\begin{align}
\Pr(\Ec_3^c)\leq 2^{2nBr}\exp(-\frac{2n\delta^2\epsilon_2^2}{B^2\rho^4})=2^{2nBr}\exp(-\frac{2n\delta^2}{B^2\rho^4})
\end{align}
Therefore, combining the bounds for the selected parameters, we obtain
\begin{align}
\Pr(\Ec_1 \cap \Ec_2 \cap \Ec_3) 
&\geq 1 - \sum_{i=1}^3 \Pr(\Ec_i^c) \\ \nonumber
&\geq 1 - 2^{4nBr} \exp\left(-\frac{2n\lambda^2\delta^2}{B^2 \rho^4}\right) - (2^{2nBr} + 1) \exp\left(-\frac{2n\delta^2}{B^2 \rho^4}\right).
\end{align}

\section{Conclusions}
\label{sec:conclusion}

In this paper, we have theoretically analyzed the performance of snapshot compressive imaging (SCI) systems under various models of binary-valued masks. While prior theoretical studies have focused on i.i.d.~Gaussian masks, such masks are rarely used in practice due to physical constraints. Instead, binary masks are commonly employed in SCI systems, and optimizing these masks is crucial for improving performance. Existing mask optimization approaches, however, are computationally intensive, provide limited theoretical insights, and are often sub-optimal due to the inherent non-convexity of the problem. To address these gaps, we have characterized the performance of SCI systems under three distinct binary mask models: fully independent masks, in-frame dependent masks, and out-of-frame dependent masks. Our analysis confirms the observations in the literature that the optimal performance for i.i.d.~binary masks is achieved when the probability of 1s is less than $0.5$. Furthermore, we establish that any form of dependence—whether in-frame or out-of-frame—deteriorates system performance, providing a novel theoretical justification for the importance of mask independence.  These findings offer a rigorous and comprehensive understanding of the role of binary masks in SCI systems, paving the way for more effective mask design and optimization. Simulation results validate our theoretical predictions and shed further light on the interplay between mask design and system performance.
\newpage
\appendix

\section{Concentration Inequalities for Dependent Random Variables}
\label{app:a}

Here, we briefly review the key results of \cite{kontorovich2008concentration} which we  use in proving Theorem \ref{thm:2}. Consider a collection of random variables $(X_i)_{1\leq i\leq n}$ taking values in a countable space $\Sc$. Assume $X_i$ are the coordinate projections defined on the probability space $(\Sc,\Fc,\Pr)$. Let $\Sc^n$ be equipped with the Hamming metric $d : \Sc^n \times \Sc^n \rightarrow [0, \infty)$, defined as  $d_{\rm H}(x, y)~\dot{=}\sum_{i=1}^n \mathbb{1}_{\{x_{i}\not=y_{i}\}}.$
Let $\E$ denote expectation with respect to $\Pr$. Also, given two random variables $Y$ and $Z$, let $\Lc(Z | Y = y)$ denote the conditional distribution of $Z$ given $Y = y$.

For the metric probability space denoted as $(\Sc^n,d_{\rm H},\Pr)$, we define the following mixing coefficients. For $1\leq i<j\leq n$, let
\begin{align}
 \Bar{\eta}_{ij}  ~\dot{=} \sup_{{y^i-1}\in\Sc^{i-1},w,\hat{w}\in\Sc} \eta_{ij}(y^{i-1},w,\hat{w}),
\end{align}
where
\begin{align}
\eta_{ij}(y^{i-1},w,\hat{w})\dot{=}\|\Lc(X^n_j|X^i=y^{i-1}w)-\Lc(X^n_j|X^i=y^{i-1}\hat{w})\|_{\rm TV}.
\end{align}
Note that $\eta_{ij}(y^{i-1},w,\hat{w})\leq 1$. 
Define  an  $n\times n$ upper-triangular matrix  $\Delta_n$ (it only considers the previous value in a sequence) such that 
\begin{align}
\textstyle    (\Delta_n)_{ij} = 
    \begin{cases}
    1, & \text{if}~i=j\\
    \Bar{\eta}_{ij}, & \text{if}~i<j\\
    0, & \text{otherwise}
    \end{cases}
\end{align}

Observe that the (usual $l_{\infty}$) operator norm of the matrix $\Delta_n$ is given explicitly by $ \| \Delta_n \|_{\infty}=\max_{1\leq i\leq n}H_{n,i}$,
where, for $1 \leq i\leq n-1$, $H_{n,i} ~\dot{=} (1+\Bar{\eta}_{i,i+1}+\cdots+\Bar{\eta}_{i,n})$.
For $i=n$, $H_{n,n}=1$  \cite{kontorovich2008concentration}. 

Using these definitions, the desired concentration result can be expressed as follows.

\begin{theorem}[Theorem 1.1 in \cite{kontorovich2008concentration}]\label{thm:app-a-1}
Suppose $\Sc$ is a countable space, $\Fc$ is the set of all subsets of $\Sc^n$, $\Pr$ is a probability measure on $(\Sc^n, \Fc )$ and $\varphi:\Sc^n\to \mathbb{R}$ is a $c$-Lipschitz function (with respect to the Hamming metric) on $\Sc^n$ for some $c > 0$. Then for any $t > 0$,
\begin{align}
\Pr\{\| \varphi - \E\varphi\|\geq t\}\leq 2\exp(-\frac{t^2}{2nc^2\|\Delta_n\|^2_\infty}).\label{eq:main-1-app1}
\end{align}

\end{theorem}

For the particular case when $(X_1, \ldots , X_n)$ is a (possibly inhomogeneous) Markov chain, the bound in Theorem \ref{thm:app-a-1} simplifies further. More precisely, given any initial probability distribution $p_0(\cdot)$ and stochastic transition kernels $p_i(\cdot|\cdot)$, $1\leq i\leq n-1$,  $1\leq i\leq n$, let the probability measure $\Pr$ on $\Sc^n$ be defined by
\begin{align}
\textstyle    \Pr\{(X_i, \ldots, X_i)=\xv\}=p_0(x_1)\prod^{i-1}_{j=1}p_j(x_{j+1}|x_j),\label{eq:Markov-def}
\end{align}
for any $1\leq i\leq n$ and any $\xv=(x_1,\ldots,x_i)\in\Sc^i$. Moreover, for $1\leq i\leq n-1$, let $\theta_i$ denote the $i$th contraction coefficient of the Markov chain defined as
\begin{align}
    \theta_i~~\dot{=} \sup_{x',x''\in\Sc}\| p_i(\cdot|x')-p_i(\cdot|x'')\|_{\rm TV}.
\end{align}
for $1\leq i\leq n-1$, and define
\begin{align}
    M_n~~\dot{=}\max_{1\leq i\leq n-1}(1+\theta_i+\theta_i\theta_{i+1}+\cdots+\theta_i\cdots\theta_{n-1}).\label{eq:def-Mn}
\end{align}
Then Theorem \ref{thm:app-a-1} can be simplified as follows. 
\begin{theorem}[Theorem 1.2 from \cite{kontorovich2008concentration}]\label{thm:app-a-2}
Suppose that $\Pr$ is the Markov measure on $\Sc^n$ described in \eqref{eq:Markov-def} and  $\varphi:\Sc^n\to \mathbb{R}$ is a $c$-Lipschitz function (with respect to the Hamming metric) on $\Sc^n$ for some $c > 0$. Then for any $t > 0$,
\begin{align}
\textstyle    &\Pr\{\| \varphi - \E\varphi \|\geq t\}
    \leq 2\exp\left(-\frac{t^2}{2nc^2M_n^2}\right),
\end{align}
where $M_n$ is defined in \eqref{eq:def-Mn}. 
\end{theorem}


\bibliographystyle{unsrt}
\bibliography{myrefs.bib}

\end{document}